\documentclass[]{spie}  

\usepackage{amsmath,amsfonts,amssymb}
\usepackage{graphicx}
\usepackage[colorlinks=true, allcolors=blue]{hyperref}
\usepackage[algoruled,lined]{algorithm2e}
\usepackage[table,xcdraw]{xcolor}
\usepackage{siunitx}

\title{Fried Parameter Estimation from Single Wavefront Sensor Image with Artificial Neural Networks}

\author[abc]{Jeffrey Smith}
\author[b]{Taisei Fujii}
\author[c]{Jesse Cranney}
\author[b]{Charles Gretton}
\affil[a]{Eindhoven Technical University, Department of Electrical Engineering, Eindhoven, Netherlands}
\affil[b]{Australian National University, School of Computing, Canberra,
Australian Capital Territory, Australia}
\affil[c]{Australian National University, Research School of Astronomy and Astrophysics, Advanced Instrumentation Technology Centre, Canberra, Australian Capital Territory, Australia}
\authorinfo{Further author information: (Send correspondence to Jeffrey Smith)\\Jeffrey Smith: E-mail: j.p.smith@tue.nl}

\begin{document} 
\maketitle

\newcommand{\Omit}[1]{}

\begin{abstract}
Atmospheric turbulence degrades the quality of astronomical observations in ground-based telescopes, leading to distorted and blurry images. Adaptive Optics (AO) systems are designed to counteract these effects, using atmospheric measurements captured by a wavefront sensor to make real-time corrections to the incoming wavefront.
The Fried parameter, $r_{0}$, characterises the strength of atmospheric turbulence and is an essential control parameter for optimising the performance of AO systems and more recently sky profiling for Free Space Optical (FSO) communication channels. 
In this paper, we develop a novel data-driven approach, adapting machine learning methods from computer vision for Fried parameter estimation from a single Shack-Hartmann or pyramid wavefront sensor image.
Using these data-driven methods, we present a detailed simulation-based evaluation of our approach using the open-source COMPASS AO simulation tool to evaluate both the Shack-Hartmann and pyramid wavefront sensors. Our evaluation is over a range of guide star magnitudes, and realistic noise, atmospheric and instrument conditions. 
Remarkably, we are able to develop a single network-based estimator that is accurate in both open and closed-loop AO configurations. Our method accurately estimates the Fried parameter from a single WFS image directly from AO telemetry to a few millimetres.
Our approach is suitable for real time control, exhibiting $0.83ms$ $r_0$ inference times on retail NVIDIA RTX 3090 GPU hardware, and thereby demonstrating a compelling economic solution for use in real-time instrument control. 

\Omit{ Atmospheric turbulence degrades the quality of astronomical observations in ground-based telescopes, leading to distorted and blurry images. Adaptive Optics (AO) systems are designed to counteract these effects, utilising the turbulence information captured by a wavefront sensor to make real-time corrections to the incoming wavefront of light. The Fried parameter ($ r_{0} $) is a highly desirable atmospheric parameter that characterises the strength of atmospheric turbulence and is essential for optimising the performance of the AO system. However, real-time estimation of $ r_{0} $ remains a challenge. In this study, we explore a novel data-driven approach using Convolutional Neural Networks (CNNs) to directly estimate $ r_{0} $ from wavefront sensor images within closed-loop environments. We leverage computer vision techniques that automate the extraction of spatial features from these images to estimate characteristics of atmospheric turbulence from the sensed wavefront. Compared to model-based approaches employed in astronomy, our data-driven approach offers a conceptually simple and practical method for real-time estimation, and is not explicitly limited by modelling assumptions. We present a detailed simulation study conducted under various atmospheric and instrumental conditions to demonstrate the robustness of the model to various challenges encountered in real-world applications. This includes open and closed-loop AO systems, variations in guide star magnitude, the impact of common sources of noise, and its applicability to both Shack-Hartmann and pyramid wavefront sensor designs. Using the open-source COMPASS simulation tool, our proof-of-concept model achieves remarkable inference speeds of $0.83ms$ while also providing $ r_{0} $ estimates with a mean absolute error of $0.2cm$. We produce an accurate $ r_{0} $ estimator which, remarkably, is able to operate in an open or closed-loop AO regime without reconfiguration. We also investigate this counter-intuitive result in depth for its academic value. These results are indicative of real-world Fried parameter estimator performance, with implications for advancing observation scheduling and AO controller optimisation.}
\end{abstract}


\section{INTRODUCTION}
\label{sec:intro}  

In recent {works~\cite{smith:2022, smith:2023, smith:2023a, Pou:24}}, Machine Learning (ML) methods have demonstrated the ability to accurately estimate the wavefront from a single Wavefront Sensor (WFS) image taken from Adaptive Optics (AO) telemetry data in simulation. In these studies, the trained neural networks were found to be robust to parameters such as the Fried parameter, implying that a simpler network may be able to extract this directly from the WFS. This would be a useful tool for astronomers, as knowledge of such parameters are necessary to correctly configure instrumentation for observations. To avoid miss-measurement, incumbent methods~\cite{wilson:2002, guesalaga:2018} require that AO systems be `opened' at regular intervals to measure the Fried parameter, interrupting observations operating in closed-loop. These interruptions typically take seconds to minutes, wasting valuable observation time. 

By applying our ML methods, we demonstrate with state-of-the-art simulation {tools~\cite{Ferreira2018COMPASS:Systems}}, a method that can directly and accurately estimate the Fried parameter from the closed-loop AO telemetry. This method would allow astronomers to measure the Fried parameter in real time, avoiding interruptions to observations. Our ML methods may also be useful for characterisation of optical channels and site profiling for Free Space Optical (FSO) ground stations.

\subsection{Adaptive Optics}
Atmospheric turbulence stands as a significant challenge to imaging in ground-based astronomical observations. While the cause of atmospheric turbulence can be attributed to various factors, the propagation of light from distant objects through layers of atmospheric turbulence creates aberrations in the wavefront. These aberrations translate to distorted, blurry images, which constrain the precision and level of detail achievable in such observations. 

Adaptive Optics (AO) systems are designed to counteract the effects of atmospheric turbulence during astronomical observations in real-time. The typical \emph{single conjugate} AO system contains three components in a closed-loop control sequence: (i) a deformable mirror that can rapidly and precisely change its shape to counteract aberrations, (ii) a wavefront sensor that collects information about the aberrated wavefront, and (iii) a real-time controller that interprets the sensor data, reconstructs the aberrations, and sends the necessary corrections to the deformable mirror. Such a ‘closed-loop’ operation of the AO system functions in real-time to correct for continuously evolving atmospheric conditions, ultimately enhancing image quality. 
\Omit{In contrast, ‘open-loop’ operations position the wavefront sensor before the deformable mirror, capturing the uncorrected wavefront information but without effective real-time correction capabilities. The choice between closed-loop and open-loop configurations depends on the design constraints and requirements of the AO system.}
Specifically, the wavefront sensor is employed to capture the shape of the incoming wavefront and translate these details into an image. The exact procedure for this translation varies with the design of the wavefront sensor. Within the scope of this thesis, we focus on the Shack-Hartmann and pyramid wavefront sensor designs.

The quality of wavefront sensor images, and therefore the performance of the AO loop, is subject to various factors. The AO system often requires a bright guide star close to the astronomical target as a reference point for measuring the wavefront distortions. The brightness of this guide star, referred to as its magnitude, is crucial to ensure that the AO sensor receives a sufficient amount of light to make accurate measurements and adjustments. AO systems are also susceptible to various sources of noise, including photon noise and readout noise, which degrade the integrity of the incoming wavefront. 

\subsection{The Fried Parameter}

One method of quantifying atmospheric turbulence is through the Fried parameter ($ r_{0} $), also known as Fried’s coherence length \cite{fried:1966}. The Fried parameter defines the average size of a circular aperture over which there is an expected wavefront error of 1 radian squared, and typically ranges from $5\ \mathrm{cm}$ in average conditions to $20\ \mathrm{cm}$ in optimal conditions. The Fried parameter is a useful environmental measurement for assessing the quality of observations that may be obtained under current atmospheric conditions, with applications in observation scheduling and site characterisation \cite{griffiths:2023}. 

Within the context of AO, the Fried parameter is especially useful in optimising the performance of the real-time controller. The \emph{model-based} tomographic algorithms used in controller optimisation are founded on statistical models of atmospheric turbulence, and their performance is dependent on the knowledge of atmospheric parameters, including the Fried parameter \cite{fusco:2001}. Therefore, a real-time estimate of such parameters is valuable for optimising overall AO performance. More recently, interest in FSO communication has driven research in measurement of sky parameters such as the Fried parameter for potential ground station locations for satellite uplink~\cite{griffiths:2023}.

Most approaches bypass the explicit calculation of the Fried parameter by estimating the atmospheric refractive index structure constant, which quantifies the strength of the turbulence as a function of height. Various methods exist to measure this characterisation of turbulence, including SCIDAR~\cite{vernin:1973} and SLODAR~\cite{wilson:2002}. However, these methods face limitations in providing real-time updates to atmospheric profiles, as processing high-dimensional data over many iterations is computationally demanding. Nonetheless, contemporary model-based approaches~\cite{guesalaga:2018, zhang2020} have achieved solutions that can run in the order of a few minutes to a few seconds. 

\subsection{Convolutional Neural Networks in Astronomy}

In recent years, Artificial Neural Networks (ANNs), a data-driven approach commonly employed in machine learning tasks, have garnered attention for their ability for adaptive non-linear feature extraction. Particularly in the realm of computer vision, Convolutional Neural Networks (CNNs) have enabled computers to automate tasks that involve the recognition of visual patterns and objects within images, and extrapolation from previously unseen data. Such technologies have facilitated transformative progress in various visual tasks across many fields, including astronomy. 

Several studies have demonstrated the power of ANNs in estimating the atmospheric refractive index structure constant from meteorological data~\cite{wang:2016}, and forecasting the profile of the near future~\cite{hou:2023}. In the realm of AO, ANNs have been instrumental in enhancing tasks such as wavefront reconstruction~\cite{dubose:2020}, wavefront estimation~\cite{liu:2020}, and image restoration~\cite{shi:2019}, showcasing their versatility. However, applying data-driven methods directly to wavefront sensor images for parameter estimation remains relatively unexplored. Recently, a new data-driven approach using conditional Generative Adversarial Networks (cGANs)~\cite{smith:2023, smith:2022} and UNets~\cite{smith:2023a, Pou:24} for accurate wavefront estimation inferred from a single WFS image, with improved computational performance and flexibility when compared to state-of-the-art model-based approaches. 
This approach excels at accurate wavefront estimation from a single wavefront sensor frame, and has shown that trained ANNs are robust to variation of Fried and other atmospheric parameters. This motivates our work, as it is this robustness that suggested an ANN may be trained to estimate the Fried parameter, and potentially others.

\subsection{Contributions and Research Goals}

In this paper we present a novel, data-driven technique for real-time estimation of the Fried parameter directly from single frames of wavefront sensor data from AO loop telemetry. Our method not only exhibits a remarkable level of accuracy but also offers a conceptually simpler alternative to existing methods, bypassing the need for explicit modelling assumptions, such as those associated with system geometry. Instead, our method derives insights directly from the available data, making it an attractive choice for practical applications. Furthermore, inference from a single WFS image and low inference time directly from the closed-loop allows for real-time estimations during observation -- unattainable with current methods. This has immediate applications in enhancing AO controller performance and optimising observation scheduling.

To achieve these outcomes, we employ a CNN to extract the spatial characteristics associated with wavefront distortions present in wavefront sensor images. Notably, we demonstrate that this approach leverages the available non-linear, higher spatial frequency information that is lost in incumbent methods~\cite{wilson:2002}. 
We investigate the feasibility and efficacy of a CNN in simulated environments that closely replicate the challenges encountered in real-world applications. Our study examines the robustness of the model under various conditions, including open and closed-loop WFS data, variations in guide star magnitude, and the presence of common sources of noise. Furthermore, we demonstrate the versatility of the approach by applying it to both Shack-Hartmann and pyramid wavefront sensors. In doing so, we find compelling evidence to suggest that our network based methods infer atmospheric turbulence from features residing in the non-linear, high spatial frequency information common to both open and closed-loop WFS measurements.

\section{Background}
\label{sec:background}

\subsection{Adaptive Optics}\label{sec:ao}
\begin{figure} [ht]
   \begin{center}
   \begin{tabular}{c} 
   \includegraphics[width=7cm]{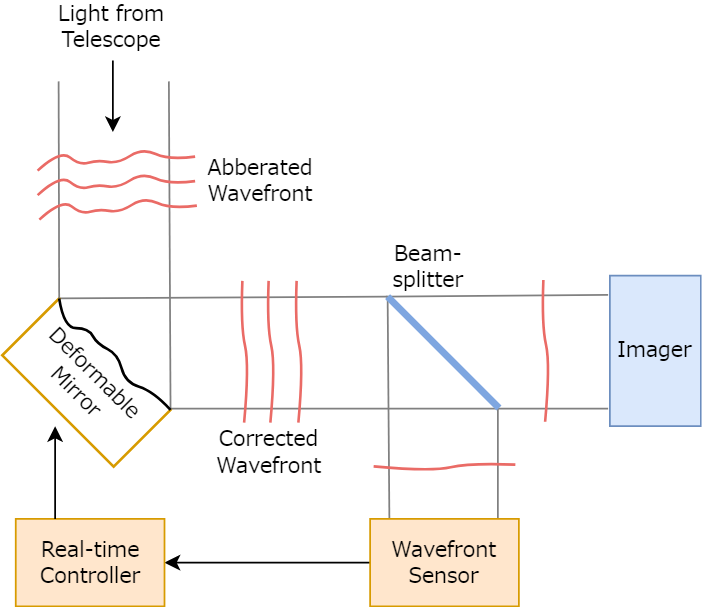}
	\end{tabular}
	\end{center}
   \caption{Diagram of a typical Adaptive Optics system. \label{fig:ao_loop} } 
   \end{figure}

AO systems are designed to counteract the effects of atmospheric turbulence during astronomical observations in real-time operation of large, optical telescopes. Initially proposed in the 1950s by \cite{babcock:1953}, the basic AO system comprises three fundamental components that operate within a closed feedback loop: a wavefront sensor (WFS), a deformable mirror, and a control system (Figure~\ref{fig:ao_loop}). The deformable mirror counteracts some of the aberrations present in the incoming wavefront, using a reflective surface that can be rapidly and precisely adjusted by a grid-like geometry of actuators. A portion of the incoming light is diverted towards the WFS, which detects and estimates residual aberrations in the wavefront. The controller then interprets the sensor’s input data and computes the necessary adjustments to the shape of the deformable mirror to minimise the distortions. The deformable mirror updates its corrections based on the controller’s feedback and continues this operation as the incoming wavefront evolves. These components collectively form a system that actively monitors and compensates for the impact of atmospheric turbulence during astronomical observations.

An AO system can be designed to operate in \textit{open-loop} or in \textit{closed-loop}. In a closed-loop system, the WFS is positioned after the deformable mirror, such that the adjustments made by the deformable mirror are continuously guided by real-time feedback from the WFS. Conversely, in an open-loop system, the WFS is situated upstream of the deformable mirror, thus bypassing the corrections applied by the deformable mirror, resulting in the wavefront sensor observing the uncorrected wavefront. 

The Shack-Hartmann Wavefront Sensor (SH-WFS) is a type of WFS that is commonly used in AO systems, designed to capture the incoming wavefront and convert it into an intensity image that encodes the local gradients of the wavefront over small sub-regions of the telescope aperture. The SH-WFS consists of an array of micro-lenses, each focusing a small portion of the incoming wavefront onto a spot on an image sensor. As such, the image captured by the SH-WFS consists of a granular map of multiple lenslet spots, as shown in Figure~\ref{fig:wfs_images}. 

\begin{figure} [ht]
   \begin{center}
   \begin{tabular}{c} 
   \includegraphics[width=0.6\linewidth]{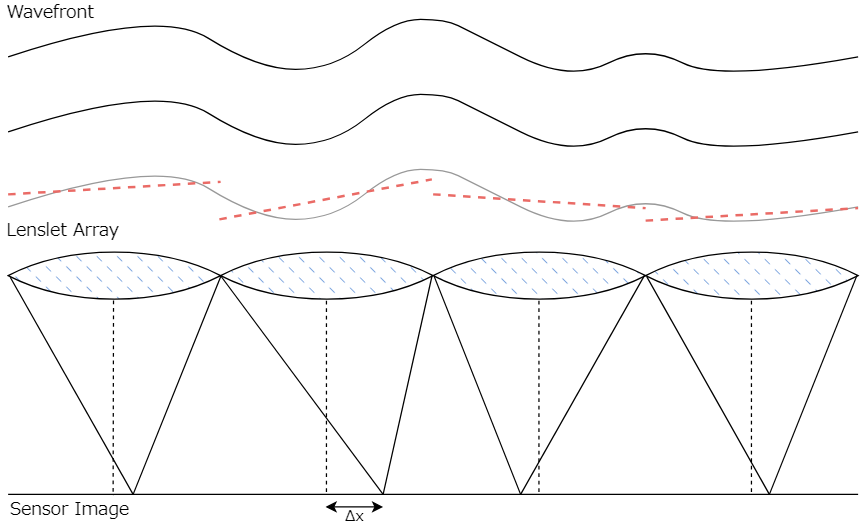}
	\end{tabular}
	\end{center}
   \caption{A simplified 1-dimensional example of the Shack-Hartmann wavefront sensor lenslets, illustrating the displacement of the focal point on the sensor due to the aberrated wavefront. This displacement $ \Delta x $ is then used to calculate the local wavefront gradient.\label{fig:shwfs_aperture}}
   \end{figure}

\begin{figure} [ht]
   \begin{center}
   \begin{tabular}{c} 
   \includegraphics[width=0.8\linewidth]{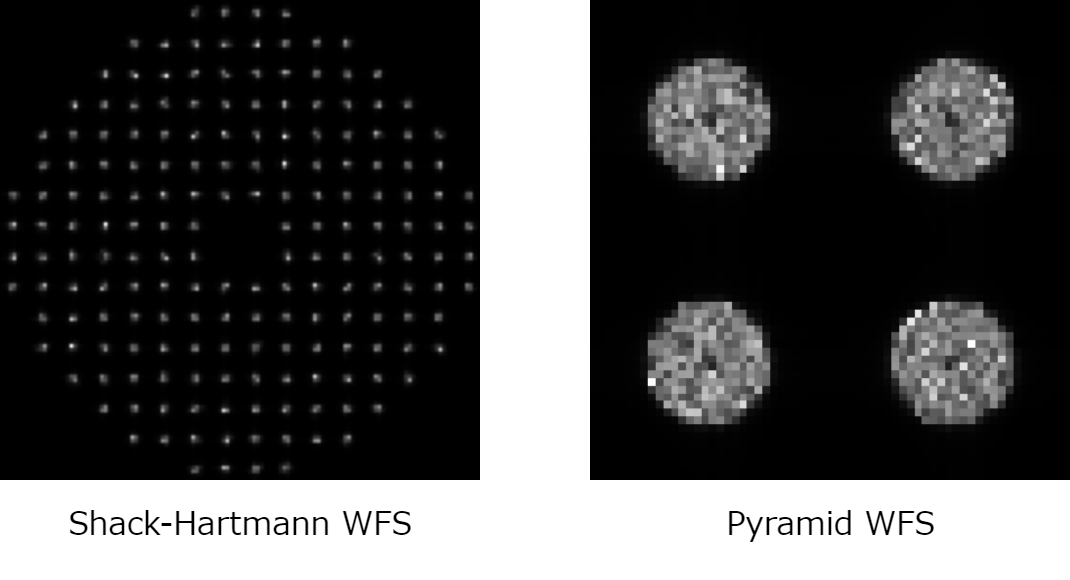}
	\end{tabular}
	\end{center}
   \caption{A comparison of images captured by the Shack-Hartmann wavefront sensor (left) and pyramid wavefront sensor (right), using the COMPASS AO simulation software.\label{fig:wfs_images}} 
   \end{figure}

The displacement of these spots from the centre of each sub-aperture is proportional to the local wavefront gradient. Figure~\ref{fig:shwfs_aperture} illustrates a simplified 1-dimensional example of the tilting of the focal spot off-axis due to the aberrated wavefront, which is subsequently estimated based on its displacement from the centre of the sub-aperture using a centroiding algorithm. The control system then computes these local wavefront gradients to estimate the global wavefront distortions over the entire sensor area. 

The sub-aperture size is a fundamental parameter in Shack-Hartmann wavefront sensors. A smaller sub-aperture size enhances the spatial sampling capability of the SH-WFS, allowing for better characterisation of fine-scale wavefront aberrations. However, the need for each sub-aperture to receive sufficient light over noise for practical wavefront analysis ultimately imposes an upper limit on achievable spatial resolution. Moreover, the SH-WFS architecture's reliance on centroids to abstract the complex wavefront into small linear components inherently results in the loss of nonlinear information at higher spatial frequencies. As sub-aperture images are simplified to points on an XY plane when interpreted by the AO loop controller, the remaining non-linear details captured by these lenslets are lost. As such, linear corrections are only applied for aberrations at the low spatial frequencies, leaving aberrations in the medium to high spatial frequencies unaccounted for. 

\begin{figure} [ht]
   \begin{center}
   \begin{tabular}{c} 
   \includegraphics[height=8cm]{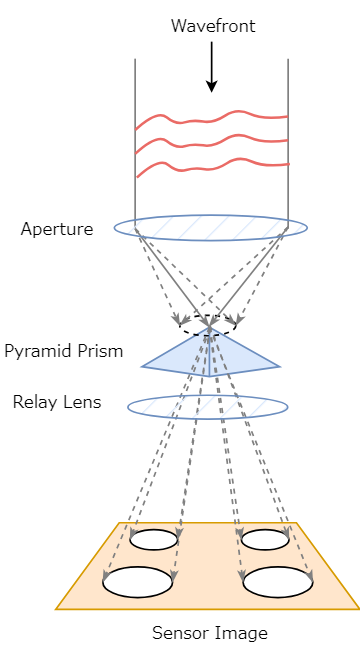}
	\end{tabular}
	\end{center}
   \caption{A typical pyramid wavefront sensor setup, illustrating the splitting of the incoming wavefront by the pyramid-shaped prism into four separate pupil images. \label{fig:pyr_wfs}}
   \end{figure}

The pyramid wavefront sensor (Py-WFS) \cite{ragazzoni:1996} is another type of wavefront sensor that is commonly used in AO systems, that employs a pyramid-shaped optical prism placed at the focal plane to split the incoming wavefront into four separate paths, each sampling a different quadrant of the focal plane. The pupil of each path is then re-imaged onto distinct regions of a sensor through a relay lens, resulting in an intensity distribution that carries information of the wavefront (Figure~\ref{fig:pyr_wfs}). In the absence of optical aberrations, these four pupil intensities would be identical. However, when aberrations are present, the intensity distribution among the four pupils on the image sensor becomes uneven. It is from these variations in pixel intensity across the pupil images that the Py-WFS measures the local wavefront gradients. To reduce the nonlinear coupling of modes, the pyramid is moved in an orbital pattern around the optical axis at a high frequency (greater than the WFS sampling frequency); a process referred to as \textit{modulation} in the Py-WFS literature. Modulations of this nature have been demonstrated to improve the linearity of the Py-WFS.
The final result captured by the Py-WFS is a mapped representation featuring four distinct pupil images arranged in a grid pattern, as shown in Figure~\ref{fig:wfs_images}.

\subsection{Turbulence Measurement for Large Telescopes}
\label{sec:fried}

The spatial resolution effects of atmospheric seeing are often quantified through the Fried parameter, also known as Fried’s coherence length \cite{fried:1966}. The Fried parameter is defined as the average size of a circular aperture over which the wavefront of light from distant objects remains relatively unperturbed by atmospheric turbulence, given by an expected wavefront error of 1 radian squared.

The Fried parameter ($ r_{0} $) for wavelength $ \lambda $ can be expressed in terms of the atmospheric refractive index structure constant $ C_{N}^{2}(h) $, which quantifies the strength of atmospheric turbulence as a function of its height $ h $ above the ground, and the angular distance of the source from zenith ($ \gamma $) :

\[ r_{0} = \left [ 0.423k^{2}(cos\gamma )^{-1}\int C_{N}^{2}(h) dh \right ]^{-\frac{3}{5}} \]

where the angular wave-number $ k = \frac{2\pi}{\lambda } $.

\begin{figure} [ht]
   \begin{center}
   \begin{tabular}{c} 
   \includegraphics[width=0.8\linewidth]{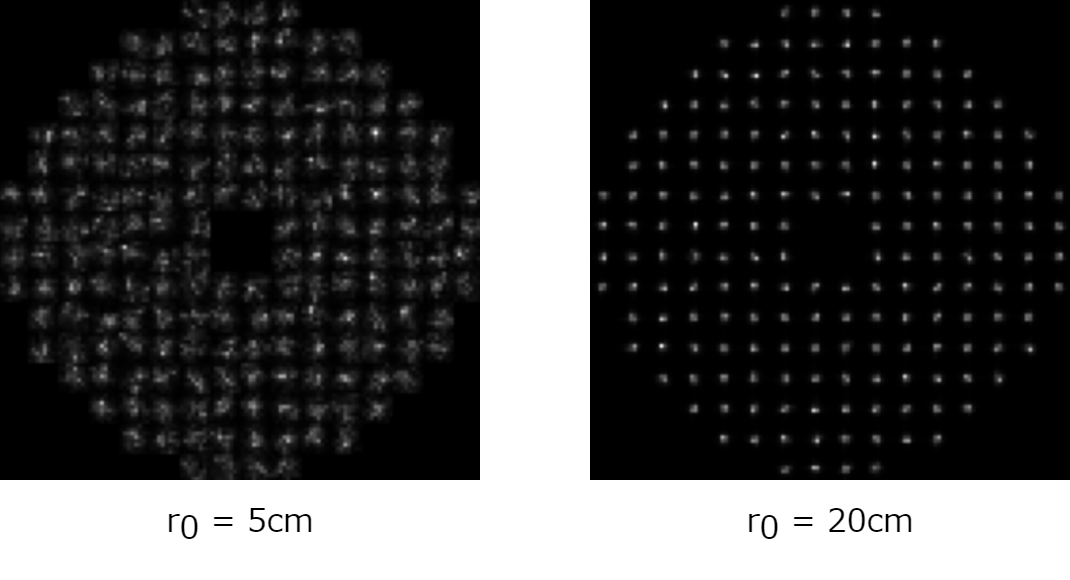}
	\end{tabular}
	\end{center}
   \caption{A comparison of Shack-Hartman wavefront sensor images taken from a telescope with a diameter of $8m$, with $ r_{0} = 5cm$  (left) and $ r_{0} = 20cm$  (right).
   \label{fig:shwfs_fried}}
   \end{figure}

The parameter is most commonly expressed in centimetres, with larger values of $ r_{0} $ associated with less severe atmospheric turbulence, and consequently, better seeing conditions. Generally, the Fried parameter varies from $ r_{0} = 20\ \mathrm{cm}$  at exceptional sites to $ r_{0} = 5\ \mathrm{cm}$ at typical sea-level sites (Figure~\ref{fig:shwfs_fried}). 

The Fried parameter is a highly desirable environmental measurement for assessing the quality of observations that can be achieved under prevailing atmospheric conditions, having implications in optimising telescope scheduling and site characterisation \cite{griffiths:2023}. 

Furthermore, knowledge of $ r_{0} $ is integral in optimising the AO controller to make accurate real-time corrections in the AO loop. The AO controller relies on tomographic algorithms using minimum mean square error (MMSE) as estimators to determine the optimal shape of the deformable mirror \cite{fusco:2001}. The goal of these algorithms is to calculate the adjustments to the shape of the deformable mirror(s) that minimises the residual wavefront error given the measurements from the wavefront sensor(s). These \emph{model-based} tomographic algorithms are dependent on statistical models of atmospheric turbulence (e.g., Kolmogorov turbulence model; \cite{kolmogrov:1941}), and in turn, rely on parameters such as the $ C_{N}^{2}(h) $ profile and the Fried parameter to make statistically optimal estimations. Given the continuously changing nature of the atmosphere, it is imperative that these parameters are updated frequently to optimise real-time corrections in the AO loop. While the exact frequency of parameter updates depends on the purpose and requirements of the telescope, real-time updates to the atmospheric profile are desirable to improve the performance of modern AO systems. 

The Fried parameter is typically determined through the assessment of the $ C_{N}^{2}(h) $ profile, a parameter intrinsically linked to $ r_{0} $ by definition. Various methodologies exist to quantify this characterisation of turbulence, with some methods dedicated to measuring the scintillation of the atmosphere (e.g., SCIDAR; \cite{vernin:1973}), while other techniques interpret measurements obtained from slope-based wavefront sensors such as the SH-WFS (e.g., SLODAR; \cite{wilson:2002}). Although existing model-based methods can provide updates to atmospheric profiles in the order of minutes and seconds, they are not currently available to astronomers in real-time with existing tools due to the large dimensionality of the data and the iterative nature of these statistical models, which require increasingly demanding computational resources as telescope systems scale up. Consequently, contemporary research on model-based methods tends to focus on optimising the computational efficiency of these algorithms, culminating in solutions with update frequencies spanning a few seconds~\cite{zhang2020}.

These turbulence characterisation techniques traditionally depend on the measurement of the raw turbulence state, necessitating the system to operate with the loop `open', that is, the AO correction of the wavefront is removed so that the true turbulence can be measured, rather than that of the corrected wavefront. However, as observations are generally carried out in closed-loop environments, the direct application of these profiling methods on the same telescope becomes infeasible. To compensate, such methods are frequently employed in conjunction with pseudo-open-loop control mechanisms that reconstruct the open-loop slopes from closed-loop slopes by using previous instructions issued to the deformable mirror \cite{ellerbroek:1998}. This facilitates the application of turbulence profiling techniques to inform parameters such as $ r_{0} $ in real-world astronomical contexts. Examples of this pseudo-open-loop slope-based strategy can be found in literature~\cite{zhang2020}.

\subsection{COMPASS Simulation Software}
COMPASS is a GPU-accelerated AO simulation tool designed to replicate all components of AO systems, particularly for the Extremely Large Telescope and other large telescopes~\cite{ferreira:2018}. COMPASS simulates the interplay of atmospheric conditions, telescope pupil, and the AO system in real-time, enabling researchers to explore and test various AO configurations within realistic environments. COMPASS also facilitates ML experimentation due the high speed of simulation that can provide notionally unlimited data for CNN training in a reasonable time frame or produce analysis data at run time for experimentation. 

In this project, COMPASS is crucial for simulating highly specific atmospheric conditions and generating WFS images that serve as training data for CNNs, used to assess network robustness and performance across different conditions. COMPASS integrates seamlessly with PyTorch via its Python interface, streamlining the connection to the CNN training framework. Example wavefront sensor images are shown in Figure~\ref{fig:wfs_images}.


\section{Methodology}
\label{sec:approach}

We design and perform an experiment to evaluate our data-driven ANN approach to accurately estimate the Fried parameter.  Our experimentation consists of a simulation study in which a typical AO system is modelled so that the Fried parameter can be directly specified in simulation parameters. Using the simulator we can produce a notionally unlimited training dataset and also a distinct experimental dataset. In this section we discuss necessary data transformations, training methods and network architecture developed and iteratively improved with experimental observations.
\Omit{
\begin{itemize}
    \item Refer to motivation (SHWFS -> Fried)
    \item Explain what a CNN \& encoder-style network architecture is
    \item Explain why this is appropriate for extracting the r0 parameter from the information the cGans and UNets were interpreting from the SH-WFS (and Py-WFS, Tomeu)
\end{itemize}
}
\subsection{Simulation Parameters} \label{sec:data_gen}
We utilised the COMPASS simulation software for two primary purposes: first, to generate the wavefront sensor images that comprised our training dataset, and second, to evaluate network performance with newly generated data.

As described in Section~\ref{sec:fried}, atmospheric seeing conditions vary between $ r_{0} $ values of $0.05\ \mathrm{m}$ in poor conditions and $0.2\ \mathrm{m}$ in optimal conditions. To train the network under a range of atmospheric turbulence conditions, we generated $10,000$ wavefront sensor images for each $ r_{0} $ value in the range [0.05, 0.06, … , 0.2] ($\mathrm{m}$), resulting in a dataset of 160,000 image-$ r_{0} $ pairs. In doing so, we expose the network to the range of atmospheric turbulence conditions typically expected for on-sky operations. 

\begin{table}[ht]
\centering
\caption{COMPASS simulation parameters \label{tab:config}}
\resizebox{0.9\textwidth}{!}{%
\begin{tabular}{|lcl|}
\hline
\multicolumn{1}{|l|}{}                                  & \multicolumn{1}{l|}{\textbf{Shack-Hartmann WFS Setup}} & \textbf{Pyramid WFS Setup} \\ \hline
\multicolumn{3}{|l|}{\textbf{Telescope Parameters}}                                \\ \hline
\multicolumn{1}{|l|}{Diameter}                & \multicolumn{2}{c|}{\SI{8}{\metre}}             \\ \hline
\multicolumn{3}{|l|}{\textbf{Atmospheric Parameters}}                              \\ \hline
\multicolumn{1}{|l|}{Number of Layers}        & \multicolumn{2}{c|}{1}             \\ \hline
\multicolumn{1}{|l|}{$r_{0}$}                & \multicolumn{2}{c|}{\SIrange{0.05}{0.2}{\metre}}      \\ \hline
\multicolumn{3}{|l|}{\textbf{Target Parameters}}                                   \\ \hline
\multicolumn{1}{|l|}{Wavelength}              & \multicolumn{2}{c|}{\SI{1.65}{\micro\metre}}     \\ \hline
\multicolumn{1}{|l|}{Guide Star Magnitude}    & \multicolumn{2}{c|}{3}             \\ \hline
\multicolumn{3}{|l|}{\textbf{WFS Parameters}}                                      \\ \hline
\multicolumn{1}{|l|}{Number of sub-apertures}           & \multicolumn{1}{l|}{$16\times16$}                             & {$16\times16$  }                    \\ \hline
\multicolumn{1}{|l|}{Number of pixels per sub-aperture} & \multicolumn{1}{l|}{$8\times8$}                               & -                          \\ \hline
\multicolumn{1}{|l|}{Wavelength}              & \multicolumn{2}{c|}{\SI{0.5}{\micro\metre}}      \\ \hline
\multicolumn{3}{|l|}{\textbf{AO Parameters}}                                       \\ \hline
\multicolumn{1}{|l|}{Loop frequency}          & \multicolumn{2}{c|}{\SI{1000}{\hertz}}       \\ \hline
\multicolumn{1}{|l|}{Delay}                   & \multicolumn{2}{c|}{2 frames}      \\ \hline
\multicolumn{1}{|l|}{Integrator Gain}         & \multicolumn{2}{c|}{0.4}           \\ \hline
\multicolumn{1}{|l|}{Noise}                   & \multicolumn{2}{c|}{-1 (no noise)} \\ \hline
\multicolumn{3}{|l|}{\textbf{DM Parameters}}                                       \\ \hline
\multicolumn{1}{|l|}{Number of DM actuators}  & \multicolumn{2}{c|}{17}            \\ \hline
\end{tabular}%
} 
\end{table}

The COMPASS simulation was configured according to the parameters displayed in Table~\ref{tab:config}, designed to replicate large telescope AO loop scenarios and that of wavefront estimation studies~\cite{smith:2022, smith:2023, smith:2023a}. Unless otherwise indicated, the COMPASS configuration used in this study is taken from COMPASS literature~\cite{ferreira:2018}. Variation in the datasets is introduced by altering the pseudo-random seed used to initialise simulated atmospheric conditions for each $ r_{0} $, ensuring the network was exposed to a variety of atmospheric scenarios during training and evaluation. Following these adjustments, the simulation ran continuously for 1000 frames to allow the changes in both seed and $ r_{0} $ to effectively influence the simulated atmosphere. Subsequently, WFS images were captured at intervals of 100 iterations, allowing atmospheric conditions to evolve between frames. Due to the difference in architecture, generated SH-WFS images were of dimension $128\times128$, while Py-WFS images were of dimension $64\times64$. A similarly sized test dataset of $160,000$ images was generated for performance evaluation, as discussed in Section~\ref{sec:metrics}.

During our experiments, we systematically adjusted several simulation configurations, including loop state (open/closed), guide star magnitude, and noise levels. These variations are discussed in Section~\ref{sec:results}, where their impact on experimental outcomes is examined.

\subsection{Data Transformations} \label{sec:data_trans}
To make the raw data from astronomical instruments and simulators compatible with our neural network, we applied a normalisation technique consistent with \cite{smith:2023}. To restrict pixel values to the range $[0,1]$, each pixel value was scaled by a constant that was 10\% greater than the maximum pixel value in the dataset, ensuring generalisability to unseen data with potentially higher pixel values. This scaling factor varied with guide star magnitude and noise levels, reflecting differences in received light.
\Omit{
\begin{figure}[ht]
  \centering
  \includegraphics[width=0.8\textwidth]{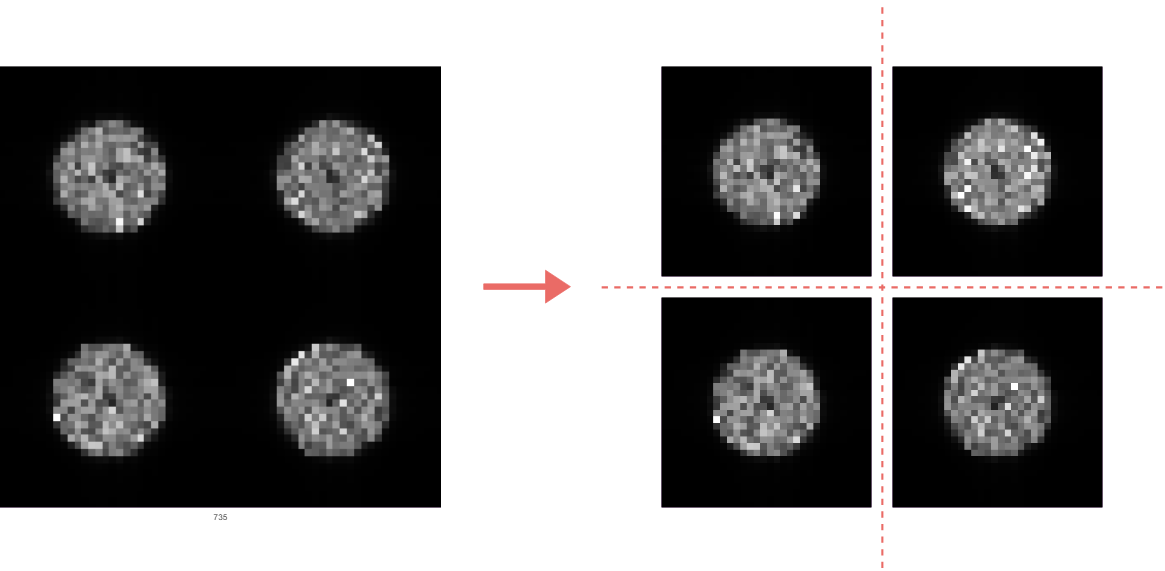}
  \caption{Division of a raw $64\times64$ image from the Py-WFS into its four separate $32\times32$ pupil images.\label{fig:pyr_split}}
\end{figure}%
}
As detailed in Section~\ref{sec:ao}, the pyramid wavefront sensor image consists of four separate lenslet images of the telescope aperture. To adapt this data for the model, we split the original $64\times64$ pixel image into four $32\times32$ pixel channels\Omit{ (Figure~\ref{fig:pyr_split})}. This necessitated slight modifications to the network architecture, discussed further in Section~\ref{sec:network_architecture}.

\subsection{Network Architecture} \label{sec:network_architecture}
We formulate the problem as a regression task that requires the model to output a numerical value as an estimate of the true $ r_{0} $. The encoder design is motivated by previous work with encoder - decoder network architectures for wavefront estimation from WFS images~\cite{smith:2022, smith:2023, Pou:24, smith:2023a}. 

We initially approached the estimation of $ r_{0} $ as a classification task, where the goal was to classify wavefront sensor images into 16 categories based on distinct $ r_{0} $ values. The early prototypes, using six convolutional layers, achieved a classification accuracy of 92-93\%, demonstrating the feasibility of a network-based approach. This success, observed in both closed and open-loop environments, motivated the shift to a regression task to improve the precision of the estimation of $ r_{0} $.

The CNN architecture for the regression task consists of:

$C16$ - $C256$ - MaxPool2d - $C512$ - $C1024$ - MaxPool2d - $C2048$ - $C4096$ - MaxPool2d - Dropout(0.3) - FC

Following the conventions of \cite{isola:2017}, $Ck$ denotes a sequential set of Convolution, Batch-norm, and ReLU layers with $k$ filters per input channel, and FC represents a fully-connected layer. The model has $72,497$ trainable parameters, with architectural parameters optimised through grid search in the benchmark experiment (Section~\ref{res:closed}).

This architecture, featuring six convolutional layers followed by a fully connected layer, was empirically found sufficient for feature extraction. Layer depth increases progressively, enabling the network to capture increasingly complex patterns across the image's sub-apertures. Zero padding at each convolutional layer ensures spatial features are preserved. Max-pooling layers \cite{zhou:1988} down-sample feature maps, enhancing robustness and enabling the model to recognise features irrespective of their spatial position.

To combat over-fitting and stabilise training, dropout \cite{sristeva:2014} and batch normalisation \cite{ioffe:2015} were incorporated. Batch normalisation was applied after each convolutional layer to normalise activations, accelerating training and improving generalisation. A dropout of $30\%$ was applied after the final convolutional layers, promoting generalisation by randomly deactivating neurons during training. The fully connected layer aggregates the features to generate the final regression output.

In experiments with both Shack-Hartmann (SH-WFS) and pyramid (Py-WFS) wavefront sensors, the architecture was adapted for the latter to handle the four-channel input from each Py-WFS lenslet image\Omit{(Figure~\ref{fig:pyr_split})}. This adjustment increased the number of kernels in the first layer to $64$, accommodating the multi-channel nature of the Py-WFS data. To mitigate the narrowing effect of stride in the network, the stride parameter was reduced, preserving spatial information more effectively, especially given the smaller dimensions of Py-WFS images. These architectural changes improved sensitivity to small-scale patterns, though at the cost of increased computational complexity.

\subsection{Training Methodology}\label{sec:training}
Each CNN was trained on a large-scale dataset of \(160,000\) wavefront sensor image-$ r_{0} $ pairs (Section~\ref{sec:data_gen}) after preprocessing steps, including resising and normalisation (Section~\ref{sec:data_trans}). The dataset was split into training (70\%), validation (20\%), and test (10\%) subsets, ensuring uniform $ r_{0} $ distributions to support model generalisation.

The model's objective was to estimate $ r_{0} $ values accurately by iteratively minimising the error between estimations and ground truth. Training proceeded over a fixed maximum of 300 epochs (\texttt{maxEpoch}), providing ample time for model convergence. The supervised training pipeline (Algorithm~\ref{algo:train}) minimised L1 loss using the Adam optimiser \cite{kingma:2015}. Training employed mini-batch gradient descent \cite{goodfellow:2016} (batch size: 256, learning rate: \(5 \times 10^{-5}\)), later adjusted to \(3 \times 10^{-5}\) for pyramid wavefront sensors (Section~\ref{res:closed}). Early stopping terminated training after five epochs of stagnant validation performance, and shuffling at each epoch to reduce over-fitting. The best model, identified by the lowest validation loss, was evaluated on the test set to assess its performance on unseen data.

\begin{algorithm}[ht]
    \caption{Training and Validation Loop \label{algo:train}}
    \KwIn{model, trainingData, validationData}
    \KwOut{best\_model}

    epoch $\leftarrow$ 0\;
    best\_model $\leftarrow$ null\;
    best\_val\_loss $\leftarrow$ +$\infty$\;

    \tcp{Iteratively train the model until maxEpoch epochs have been reached}
    \While{epoch \textless maxEpochs}{
        \tcp{Shuffle the training data}
        shuffle(trainingData)\;
        
        \tcp{Train the model over mini-batches of the training data}
        \ForEach{batch in trainingData}{
            \tcp{Predict the label based on the input images}
            predictions $\leftarrow$ model.predict(batch.images)\;
            
            \tcp{Compute the total loss over the mini-batch}
            loss $\leftarrow$ loss\_func(predictions, batch.labels)\;
            
            \tcp{Update the model parameter with respect to loss using the optimiser}
            optimiser(model, loss)\;
        }
        
        \tcp{Evaluate the model against the validation set}
        predictions $\leftarrow$ model.predict(validationData.images)\;
        validation\_loss $\leftarrow$ loss\_func(predictions, validationData.labels)\;
        
        \tcp{Update the best model based on validation loss}
        \If{validation\_loss \textless  best\_val\_loss}{
            best\_val\_loss $\leftarrow$ validation\_loss\;
            best\_model $\leftarrow$ model\;
        }
        
        \If{validation\_loss has not improved for five consecutive epochs}{
            terminate\;
        }
        
        epoch $\leftarrow$ epoch + 1\;
    }
    
\end{algorithm}

\section{Results and Evaluation}

\label{sec:results}

\subsection{Evaluation Metrics} \label{sec:metrics}

Trained models were evaluated on several metrics.  The performance of the model was tested using an unseen dataset of $160,000$ samples of simulation data generated using the same process detailed in Section~\ref{sec:data_gen}. While a test set was reserved during training, as described in Section~\ref{sec:training}, the wavefront sensor images in this test set were generated from the same set of seeds as the training set. Thus, although this test set provides an indication of the model’s generalisability to similar atmospheres, an independently generated dataset was essential to assess the model’s robustness across diverse atmospheres created from various seeds.

In evaluating the model's performance on this extensive dataset, we employed two key metrics. The first is the Mean Absolute Error (MAE), serves as a direct reflection of the model's accuracy in estimating the Fried parameter. Complementing MAE, we employed root mean squared error (RMSE), providing a deeper understanding of the spread and dispersion of errors. Models were also evaluated based on their inference speed to assess the feasibility of leveraging data-driven approaches as real-time estimators of the Fried parameter, especially when compared to contemporary model-based methods. Speed of inference measurements were taken using CUDA synchronisation and averaged over 300 repetitions. 

\subsection{Hardware and Experimental Method} 
The methodology outlined in the previous sections was implemented using COMPASS with NVIDIA CUDA Toolkit v11.2 to generate the wavefront data, and PyTorch v1.8.0 to build and train the networks. All experiments were conducted on a server with a single AMD Ryzen Threadripper 3990X (64 cores, 128 threads; 2.9 GHz clock speed), a single NVIDIA RTX 3090 24GB, and 128GB of DDR4 OC RAM.

\subsection{Results - Shack-Hartman Wavefront Sensor}

\subsubsection{Open-loop Estimator}\label{res:open}

\begin{figure}[ht]
\includegraphics[width=0.95\textwidth]{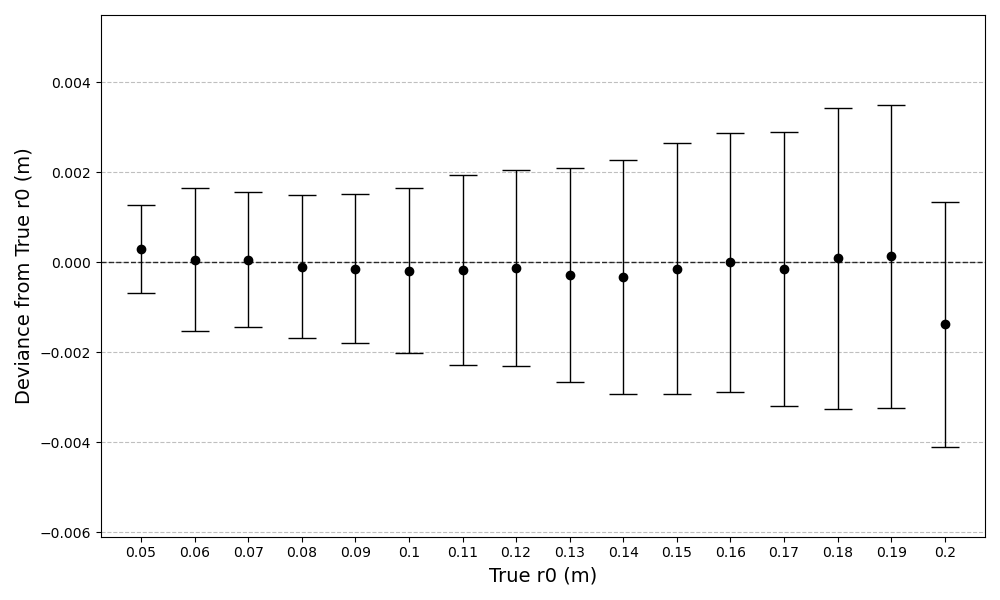}
\caption{\textbf{Open-loop network}. The mean deviance of r0 estimates from the true r0 value when the open-loop model is evaluated on \textbf{open-loop} data. A positive deviance value signifies an overestimation of the true r0. The error bars on the graph indicate one standard deviation.}
\label{fig:Open_Loop_Eval}
\end{figure}

Here we investigate the feasibility of the data-driven approach in extracting the spatial features associated with atmospheric turbulence from SH-WFS images to estimate the Fried parameter from open-loop data. We aimed to accomplish this by providing the simplest dataset to form an initial assessment of the model. 
Here, the incoming wavefront captured by the wavefront sensor is not corrected by the deformable mirror, which is required by incumbent methods to measure the Fried parameter. In closed-loop systems, incumbent methods require the loop to be `opened' for measurement of the Fried parameter to avoid altering measurement, interrupting observations for a significant period.
%
%
%
Noise was deliberately omitted from this dataset, and a bright guide star of magnitude 3 was chosen to provide optimal conditions for SH-WFS image quality.
The model was trained and evaluated on a dataset of open-loop wavefront sensor images generated as described in Section 3.2. The results are summarized in Figure~\ref{fig:Open_Loop_Eval}.

The results reveal an overall MAE of 0.00184 (0.184cm) and RMSE of 0.00242 (0.242cm), signifying that the model was able to estimate the Fried parameter directly from single wavefront sensor images with high accuracy. Approximately 95.04\% and 99.95\% of estimates fell within error margins of 0.005m and 0.01m from the true r0, respectively.
%
%
The result demonstrates real-time inference capability, with an inference time of 0.83 milliseconds on modest GPU hardware, highlighting its potential as a real-time estimator that infers the Fried parameter from a single wavefront measurement. This result eliminates the need for statistical convergence over many seconds or minutes of wavefront measurements required by incumbent methods. 

The inference speed for the model remains relatively consistent throughout the subsequent experiments. This stability can be attributed to the uniformity of model architecture across all SH-WFS experiments despite changes in training and evaluation data. 

The mean deviation of estimates from the true $ r_{0} $ remains relatively consistent across various $ r_{0} $ values, with one exception: $ r_{0} = 0.2 $, for which the model consistently underestimates. Additionally, the model exhibits a positive correlation between estimate variance and $ r_{0} $, with lower variance at the shorter $ r_{0} $ values. This observation implies that more information is available to the network in environments with poorer seeing conditions. We hypothesise that this phenomenon arises due to photon scattering within each sub-aperture of the SH-WFS sensor data caused by turbulence, causing more pixels to become illuminated in more turbulent conditions (Figure~\ref{fig:shwfs_fried}). Thus, the model is provided with additional data points to infer the Fried parameter, resulting in more stable estimation for smaller $ r_{0} $ values. This is likely influenced by SH-WFS design such as number of sub-apertures and pixels per sub-aperture. 

However, within the context of AO correction, this is counteracted by the diminishing returns of $ r_{0} $ accuracy in better seeing conditions. In practice, the accuracy of corrections made by the AO controller becomes less and less critical with longer $ r_{0} $ values, as seeing conditions are already relatively good at such $ r_{0} $ values. Therefore, this issue would not be significant in practical applications. 

\subsubsection{Closed-loop Estimator}\label{res:closed}
\Omit{
\begin{itemize}
    \item 4.2.1. Open \& closed-loop Estimators \label{res:closed}
    \item 4.2.2. Synthesis of Open \& Closed-loop Estimators
    \item 4.2.3. Cross-Evaluation of Open \& Closed Estimators
    \item 4.2.4. Variations in Guide Star Magnitude
    \item 4.2.5. Variations in Noise
    \item 4.2.6. Variations in Guide Star Magnitude with Noise
    \item 4.2.7. Application of Moving Averages
    \item 4.2.8. Extending the Application to Py-WFS
\end{itemize}

}

Building upon the results of our initial experiment, we proceeded to assess the effectiveness of our data-driven approach in closed-loop environments. The ability to estimate the Fried parameter in closed-loop scenarios is of significant practical importance, as most astronomical observations are conducted in closed-loop and would remove the need to rely on pseudo-open-loop techniques. 

As discussed previously, the wavefront captured by the WFS has already been corrected by the AO loop in closed-loop operations. During this correction process, the features in the low spatial frequencies are corrected out by the linear controller, altering information about atmospheric turbulence. This, intuitively, might pose a greater challenge to the model in quantifying turbulence compared to uncorrected WFS images. Furthermore, the residual errors introduced from this correction may also introduce some degree of randomness into the spatial features, potentially undermining the authenticity of the information compared to open-loop estimations.

\begin{figure}[ht]
\includegraphics[width=0.95\textwidth]{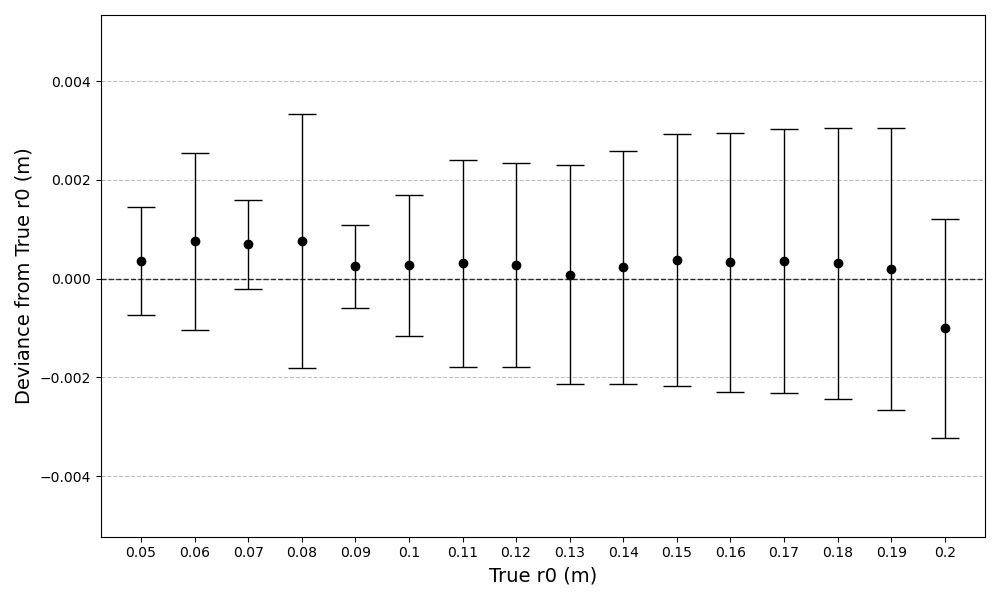}
\caption{\textbf{Closed-loop network}. The mean deviance of r0 estimates from the true r0 value when the closed-loop model is evaluated on \textbf{closed-loop} data. A positive deviance value signifies an overestimation of the true r0. The error bars on the graph indicate one standard deviation.}
\label{fig:closed2c}
\end{figure}

For this experiment, we trained and evaluated the model using a closed-loop variant of the dataset employed in the open-loop experiment. The outcomes are summarised in Figure~\ref{fig:closed2c}.

The results reveal that the model exhibited consistent and reliable estimations of the Fried parameter in closed-loop environments, with an MAE of 0.00166 ($0.166cm$) and RMSE of 0.00222 ($0.222cm$). Approximately 96.4\% and 99.99\% of estimations fell within error margins of $0.005m$ and $0.01m$, respectively, compared to the true $ r_{0} $ values. 

Contrary to our initial hypothesis, we found that the closed-loop model achieved performance similar to that of the open-loop model. This suggests that information pertaining to atmospheric turbulence is still available in corrected wavefront data.
These findings align with those of \cite{smith:2023a}, that demonstrated the wavefront sensors retained higher-order non-linear spatial information above the cut-off frequency of the deformable mirror in an AO system.
This outcome holds significant implications for real-world applications, as a real-time Fried parameter estimator would be particularly practical in the closed-loop environments commonly used for astronomical observations.

We propose a plausible explanation for the consistent model performance between the open and closed-loop scenarios. It is likely that the features captured by the model remain relatively unaffected by the AO controller. During this correction process, the features in the low spatial frequencies are detected by the centroiding algorithm and corrected out by the linear controller. However, this also implies that the information in the higher spatial frequencies that are undetected and uncorrected by the AO loop is preserved between both scenarios. It is reasonable to conclude that the high spatial frequency information (beyond centroids) is being used in both sensor modes (open and closed-loop) to estimate the Fried parameter, as this is the only facet of turbulence statistics that is common to both regimes. 

Estimation means are uniform across $ r_{0} $ values at slightly overestimating the true $ r_{0} $, with the exception of the outlier, $ r_{0} = 0.2 $, which again is underestimated like in previous open-loop results. With the exception of a few outliers at the shorter $ r_{0} $ values, this model also adheres to the previously outlined trend of increasing variance for longer $ r_{0} $ values. This observation suggests that this relationship is not necessarily restricted to open or closed-loop environments, but may be a product of the SH-WFS architecture or the task itself.

\subsubsection{Combined Open and Closed-loop Estimation}\label{openclosed}

We now further investigate our previous hypothesis regarding the model’s use of information within the spatial frequencies beyond the linear controller to estimate the Fried parameter. We conducted an experiment where the network was trained on both of the previously established training sets, encompassing both open and closed-loop data, and evaluating its performance on both corresponding test sets. Ordinarily, one would anticipate that the network's performance might become unstable under such circumstances, given the model needs to learn a strategy that handles two different types of training data with the same labels. 

\begin{figure}[ht]
\includegraphics[width=0.95\textwidth]{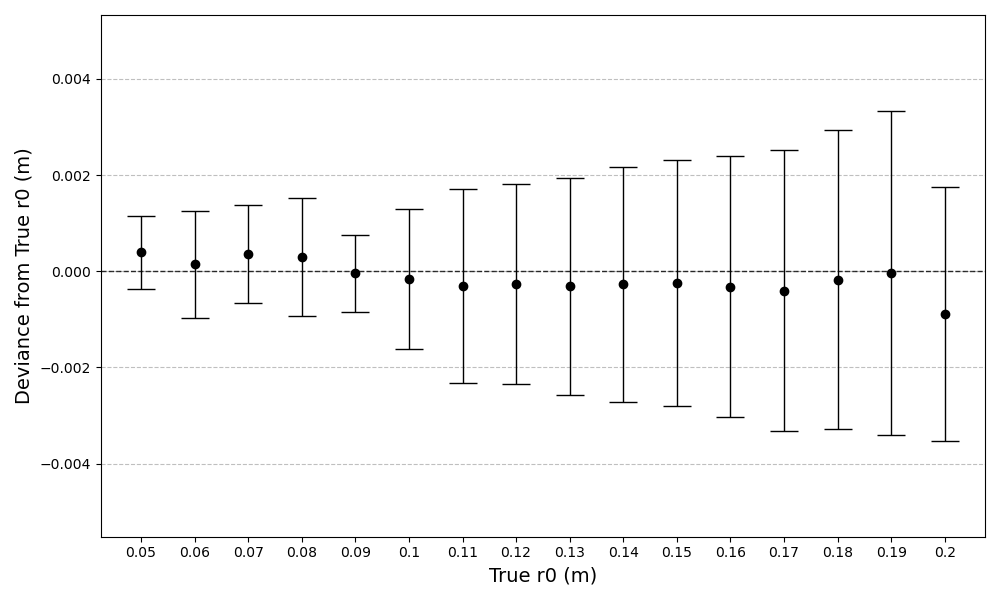}
\caption{\textbf{Combined open and closed-loop network}. The mean deviance of r0 estimations from the true r0 value with evaluation on \textbf{open-loop} data. A positive deviance value signifies an overestimation of the true r0. The error bars on the graph indicate one standard deviation.}
\label{fig:both2o}
\end{figure}

\begin{figure}[ht]
\includegraphics[width=0.95\textwidth]{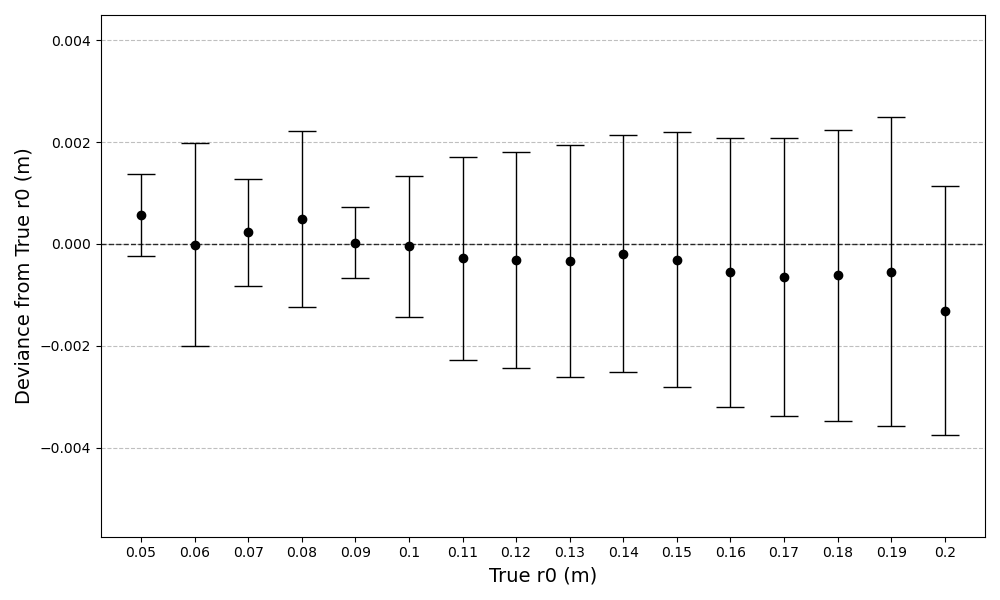}
\caption{\textbf{Combined open and closed-loop network}.The mean deviance of r0 estimations from the true r0 value with evaluation on \textbf{closed-loop} data. A positive deviance value signifies an overestimation of
the true r0. The error bars on the graph indicate one standard deviation.}
\label{fig:both2c}
\end{figure}

However, if the model manages to sustain its performance and learns to estimate the Fried parameter of both open and closed-loop wavefront sensor images simultaneously, it would strongly suggest that the model draws upon information mutual to both types of data to estimate the $ r_{0} $. This, in turn, would support the existence of turbulence information at uncorrected spatial frequencies that are preserved from open to closed-loop environments. The results are summarised in Table~\ref{tab:sh_oc}, with individual results for open and closed-loop evaluation in Figure~\ref{fig:both2o} and Figure~\ref{fig:both2c}, respectively.

Interestingly, the results reveal that the network performed equally well when evaluated on open and closed-loop data, yielding a common MAE of 0.00159. This result surpasses the performance of the two previous networks in both metrics.

The network performance shows no visible signs of instability, lending strong support to the hypothesis that it effectively leverages information common to both types of images. This evidence supports the idea that the features used to infer the Fried parameter reside in the high spatial frequencies, as low spatial frequency features are corrected out in closed-loop environments. This evidence, in turn, suggests that the network not only learns to eliminate the lower-order data present in open-loop images, but is also able to discern and utilise finer details beyond merely local gradients across the lenslets to infer turbulence. This finding aligns with the observations made by~\cite{smith:2022, smith:2023, smith:2023a}, that found that neural networks interpreted high-order non-linear spatial information from the WFS to reconstruct features that were more complex than the local gradient.

\begin{table}[ht]
\caption{Summary of MAE and RMSE scores of open-loop, closed-loop and combined networks against open and closed-loop SH-WFS data across all $ r_{0} $ values. }
\label{tab:sh_oc}
\centering
\begin{tabular}{ll|ll}
               &      & \multicolumn{2}{l}{\textbf{Evaluation Dataset}} \\
\textbf{Network} & (m)  & Open-loop             & Closed-loop             \\ \hline
Open-loop      & MAE  & 0.00184               & 0.00292                 \\
               & RMSE & 0.00242               & 0.00373                 \\ \hline
Closed-loop    & MAE  & 0.00682               & 0.00175                 \\
               & RMSE & 0.00906               & 0.00230                 \\ \hline
Combined       & MAE  & 0.00159               & 0.00159                 \\
               & RMSE & 0.00222               & 0.00221               
\end{tabular}
\end{table}

Notably, this network achieves better performance on the open-loop evaluation dataset in comparison to the network trained exclusively on open-loop data. This difference in performance could imply that compelling the network to train on both sets of data forced it to seek informative features common to both, effectively filtering out the unnecessary low-order information present in open-loop images. Alternatively, it could be attributed to the larger volume of training data in this experiment, albeit with the added challenge of handling two different data types.

Furthermore, the distribution of estimates for both open (Figure~\ref{fig:both2o}) and closed-loop (Figure~\ref{fig:both2c}) evaluation sets closely resembles the distributions observed in the dedicated open and closed-loop experiments (Sections ~\ref{res:open} and ~\ref{res:closed}). This reinforces the notion that all networks estimate the $ r_{0} $ based on similar spatial features regardless of whether it is open or closed-loop. 

In summary, this experiment provides compelling evidence that the network can effectively harness shared information from both open and closed-loop data to estimate the Fried parameter. However, further research is warranted to elucidate the exact nature of the turbulence information that is being preserved during the correction by the AO loop, and how it is then being used by the network for inference.

\Omit{
\subsubsection{Cross evaluation of Open and closed estimators (probably should cut this)}

To further substantiate the findings from our previous experiment, we conducted a cross-evaluation between the open and closed-loop networks and their evaluation sets. Specifically, we assessed the performance of the network trained on open-loop data (Section~\ref{res:open}) when evaluated on the closed-loop test set, and compared it to its baseline performance on open-loop data. Similarly, we conducted the reverse of this experiment with the network trained on closed-loop data (Section~\ref{res:closed}). Through this, we aimed to investigate the extent to which networks tended to learn and rely on the features present in both image types and, more generally, whether network performance was transferable across open and closed-loop images. 

To ensure uniformity in the networks' training and evaluation, both were retrained using a common normalisation factor used to scale pixel intensity in WFS image preprocessing. This step aimed to enable the networks to account for the full range of pixel values observed in both the open and closed-loop datasets. The results are summarised in Table~\ref{tab:cross_res}. 

The results reveal that the MAE performance of the open-loop network experienced only a slight deterioration when evaluated on closed-loop data compared to open-loop data. However, a notable increase in variance can be observed through the RMSE metric and in Figure~\ref{fig:open2c}. This evidence suggests that the network extracts features from open-loop data that are also present in closed-loop data. The heightened variance indicates that such estimates are not consistently stable. This might imply that the network is also making inferences based on features unique to open-loop data, which could potentially be misleading when inferring from closed-loop images.

In contrast, the closed-loop network exhibits a more pronounced deterioration in performance of around $0.5cm$ for both metrics when evaluated on open-loop data compared to closed-loop data (Figure~\ref{fig:closed2o}). This indicates that the features captured in closed-loop images cannot be directly applied to open-loop images.

This larger error observed in the inverse experiment could be attributed to the closed-loop network suddenly encountering previously unseen low-order features that it had not been trained on. In contrast with the experiment described in Section~\ref{res:open+closed}, this network did not have the opportunity to learn to `ignore’ the information in the lower spatial frequencies present in open-loop images. Therefore, we theorise that the network's performance deteriorates due to its inability to interpret the patterns learnt in closed-loop images in the presence of low-order information. We posit that this deteriorating effect is more pronounced in this case compared to the loss of learned spatial features in the open-loop network.

In summary, these results suggest that independently trained open and closed-loop networks tend to leverage features common to both image types.  However, there is also evidence to suggest that some features may be exclusive to each image type. Moreover, the transferability of network performance from closed to open-loop images is not as effective, while the reverse appears to be more plausible.

}
\Omit{
\subsubsection{Robustness to variation photon flux}

As discussed in Section 2.1.2, guide star magnitude directly influences the quantity of incident light available to the wavefront sensor. Different observations may have different guide star magnitudes, and the magnitude of laser guide stars is known to vary throughout the course of an observation. Consequently, it is essential to determine how variations in guide star magnitude could influence network performance and investigate model robustness to such changes.

In order to investigate this, our experiment involved two models: a reference model and an experimental model. The reference model was trained using a dataset of closedloop wavefront sensor images generated with a guide star of magnitude 9. In contrast, the experimental model was trained on a dataset of equal size, consisting of closed-loop wavefront sensor images generated with guide star magnitudes in [9, 10, 11]. The number of images for each magnitude in the experimental dataset was evenly distributed. Each model was subsequently evaluated using test sets generated with guide star magnitudes spanning [9, 10, 11, 12, 13]. By evaluating the performance of both models on images with mutually unseen guide star magnitudes of 12 and 13, we aimed to compare the models’ abilities to generalise to unseen magnitudes. The results are summarised in Table ?? and Figure ??

\begin{table}[ht]
\caption{Summary of MAE and RMSE scores of two models trained on magnitude
9 (reference) and 9-11 (experimental) datasets, evaluated across closed-loop
wavefront sensor images of magnitudes 9-13, and across all r0 values. Guide
star magnitudes unseen during training are highlighted.
\label{tab:SH_GSM_oc}}
\centering
\begin{tabular}{ll|lllll}
               &      & \multicolumn{5}{c}{\textbf{Guide Star Magnitude}} \\
\textbf{Model} & (m)  & 9        & 10       & 11      & 12      & 13      \\ \hline
Reference      & MAE  & 0.00176  & \cellcolor{gray}0.05445  & \cellcolor{gray}0.07756 & \cellcolor{gray}0.06807 & \cellcolor{gray}0.05725 \\
               & RMSE & 0.00235  & \cellcolor{gray}0.05925  & \cellcolor{gray}0.08974 & \cellcolor{gray}0.08163 & \cellcolor{gray}0.06978 \\ \hline
Without Noise  & MAE  & 0.00209  & 0.00209  & 0.00217 & \cellcolor{gray}0.02183 & \cellcolor{gray}0.05380 \\
               & RMSE & 0.00270  & 0.00271  & 0.00282 & \cellcolor{gray}0.02337 & \cellcolor{gray}0.06343 \\ \hline
With Noise     & MAE  & 0.00227  & 0.00218  & 0.00248 & \cellcolor{gray}0.01540 & \cellcolor{gray}0.03382 \\
               & RMSE & 0.00289  & 0.00278  & 0.00319 & \cellcolor{gray}0.02010 & \cellcolor{gray}0.04182
\end{tabular}
\end{table}

Initially, the reference model exhibits better performance when tested on the same magnitude it was trained on, i.e., magnitude 9. However, its performance significantly deteriorates when confronted with unseen magnitudes. Notably, this decline in performance did not exhibit a direct correlation with the magnitude of the unseen guide star. 

On the other hand, the experimental model presents a different performance pattern, initially performing worse on magnitude 9 data compared to the reference model. This divergence could be attributed to the model having only one-third of the training data dedicated to magnitude 9, resulting in limited exposure to each individual magnitude. Nevertheless, the experimental model consistently maintained accuracy across the trained magnitudes, ranging from 9 to 11, albeit at slightly higher error levels in both metrics.

The overall increase in error observed across the seen magnitudes for the experimental model, although slight, suggests that handling a broader range of magnitudes inherently presents a more challenging task. This implies the potential need for larger model architectures or the consideration of larger training set sizes for each magnitude to enhance model performance.

Despite these variances, similar to the reference model, the experimental model’s performance deteriorates when tested on unseen magnitudes. This observed increase in error across the experiments hints at some degree of overlap or correlation between the Fried parameter and guide star magnitude. This suggests that these two variables may not be entirely independent, which can subsequently result in increased error when the model is tested on unseen magnitudes.

However, intriguingly, the experimental model showcased greater robustness when dealing with unseen magnitudes that were not substantially different. For example, the error of the experimental model, when tested on magnitude 12, was noticeably lower than the error observed when testing against other unseen magnitudes, particularly when compared to the reference model.

These findings suggest that when a model is trained on an expected range of guide star magnitudes, it can exhibit robustness when handling guide star magnitudes falling within that predefined range. Furthermore, they hint at the potential for improved performance when dealing with unseen magnitudes situated in proximity to the trained range. This has practical implications, especially in astronomical observations where guide star magnitudes can vary due to factors such as changes in the brightness of the sodium layer. Adjusting for these brightness fluctuations may be feasible by defining an effective range and training the model on guide star magnitudes falling within that designated range.

\subsubsection{Robustness to noise}

Next, we sought to investigate the influence of noise on model performance. As discussed in Section 2.1.2, read-out noise and shot noise are two fundamental sources of noise that hold the potential to significantly compromise the integrity of wavefront sensor images. Given the intrinsic nature of these noise sources, it is essential to gauge the effects of these noise sources on the model’s performance and measure its robustness.

Three types of noise configurations will be investigated: (i) no noise, (ii) photon noise, and (iii) a combination of photon noise as well as a readout noise of 1 photon/pixel, reflecting a more realistic representation of noisy wavefront sensor images. Consistent with our previous experimental approach, two models were trained to investigate noiseinduced variations in model performance. The reference model was trained on a dataset of noise-free closed-loop wavefront sensor data, while the experimental model was trained on an equivalently sized dataset consisting of images encompassing all three aforementioned noise configurations. The number of images for each noise configuration within the second dataset was evenly distributed. Each model was subsequently evaluated using test sets with each of the three different noise configurations. Notably, we conducted all our experiments under a reduced guide star magnitude of 12, as the effects of noise are generally more pronounced at lower guide star magnitudes. The results are summarised in Table 4.4 and Figure 4.4.

As it can be seen, the reference model exhibited superior performance when tested on the same noise-free configuration it was trained on. However, its performance experiences a substantial decline when subjected to unseen noise configurations. Notably, this degradation in performance was correlated with the complexity of the introduced noise types; as noise levels increased in complexity, both error metrics showed corresponding increments.

In contrast, the experimental model initially performed less effectively on noise-free data. This performance gap may arise from two factors: the model’s limited exposure to noise-free instances during training or the heightened task complexity introduced by noise. However, this model showcased remarkable consistency when evaluated across the seen noise configurations, maintaining its accuracy even in the presence of varied noise. Although a decline in performance for both metrics was evident with the introduction of more complex noise, this decrease in accuracy was significantly milder when compared to the reference noise-free model. This observation suggests that while there may be some correlation between the Fried parameter and noise, the model exhibits the capacity to be trained to account for and mitigate the impacts of such noise. This implies that models can be trained on a spectrum of expected noise configurations to effectively handle the levels of noise encountered in practical settings. Such adaptability proves particularly significant considering the inevitable presence of noise in real-world observations which varies naturally depending on the instrument and observational context.
}

\subsubsection{Robustness to Photon Flux and Noise}{\label{Robust}}

Measurement of the Fried parameter is done in a variety of conditions where the ANN would ideally adapt without the need for specifically trained networks that match conditions. One commonly variable measurement parameter is the guide star magnitude or photon flux, where the number of photons incident on the WFS varies with the target. Another common effect is noise, both from photon noise due to the statistical limit where measurements show stochastic variations, and read-out noise of sensors.

Following the robustness analysis from studies of wavefront {estimation~\cite{smith:2023a}}, networks were first tested individually for robustness to photon flux and noise. Both analysis show strong robustness to both of these parameters. By combining training datasets for noise and variations in photon flux, we demonstrate here that the ANN can be trained to be robust to variations in photon flux, represented here by variation of guide star magnitude, and also to noise with the addition of shot noise and read out noise by adding both of these to the training and testing data via the COMPASS simulator.  

First as a reference we trained a network on closed-loop WFS images with a guide star magnitude 9. We then trained a single network on a dataset consisting of closed-loop WFS images with guide star magnitudes uniformly distributed in [9, 10, 11], without any noise. We then evaluated the network against a test set of closed-loop wavefront sensor images generated with guide star magnitudes spanning [9, 10, 11, 12, 13] and contrast these results in Table~\ref{tab:SH_GSM_c}.
We then train a similar network on all training and evaluation datasets were generated with both shot noise and a readout noise of 1 photon/pixel, selected as per the analysis in wavefront estimation {literature~\cite{smith:2023a}}. At these values of noise and photon flux, the guide star magnitude of 12 causes significant degradation of the simulated open-loop, and at magnitude 13 the simulated AO loop becomes unstable as a result of the noise impacting the WFS measurements. 

\begin{table}[ht]
\caption{Summary of MAE and RMSE scores of two networks trained on magnitude
9 (reference) and 9-11 (with and without noise) datasets, evaluated across closed-loop
wavefront sensor images of magnitudes 9-13, and across all r0 values. Guide
star magnitudes unseen during training are highlighted.
\label{tab:SH_GSM_c}}
\centering
\begin{tabular}{ll|lllll}
               &      & \multicolumn{5}{c}{\textbf{Guide Star Magnitude}} \\
\textbf{Network} & (m)  & 9        & 10       & 11      & 12      & 13      \\ \hline
Reference      & MAE  & 0.00176  & \cellcolor{lightgray}0.05445  & \cellcolor{lightgray}0.07756 & \cellcolor{lightgray}0.06807 & \cellcolor{lightgray}0.05725 \\
               & RMSE & 0.00235  & \cellcolor{lightgray}0.05925  & \cellcolor{lightgray}0.08974 & \cellcolor{lightgray}0.08163 & \cellcolor{lightgray}0.06978 \\ \hline
Without Noise  & MAE  & 0.00209  & 0.00209  & 0.00217 & \cellcolor{lightgray}0.02183 & \cellcolor{lightgray}0.05380 \\
               & RMSE & 0.00270  & 0.00271  & 0.00282 & \cellcolor{lightgray}0.02337 & \cellcolor{lightgray}0.06343 \\ \hline
With Noise     & MAE  & 0.00227  & 0.00218  & 0.00248 & \cellcolor{lightgray}0.01540 & \cellcolor{lightgray}0.03382 \\
               & RMSE & 0.00289  & 0.00278  & 0.00319 & \cellcolor{lightgray}0.02010 & \cellcolor{lightgray}0.04182
\end{tabular}
\end{table}

From the results in {Table~\ref{tab:SH_GSM_c}}, the reference network can accurately estimate the Fried parameter for guide star magnitude of 9, but becomes inaccurate at other magnitudes. The networks trained with several guide star magnitudes in their training data are robust to variations in the guide star magnitude and accurately infer the Fried parameter. At guide star magnitude 12, the estimates are impacted but still below that of the Reference network. 
The results for the network trained and tested with noise show similar accuracy to the network trained and evaluated without noise. This indicates that an ANN can be trained to be robust to both photon flux and noise, with only a small loss of accuracy.

\subsubsection{Applying a Moving Average to Estimation}
In practical applications for AO, networks may employ a buffering or sliding window approach with telemetry data to stabilise their output. To assess the benefits of this approach in real-time Fried parameter estimation, we sought to evaluate the stability of model inference when applied over a moving average of predictions. By quantifying the variance in these moving-average estimates, we assess the consistency and reliability of the network’s estimates over time.

\begin{figure}[ht]
\includegraphics[width=0.95\textwidth]{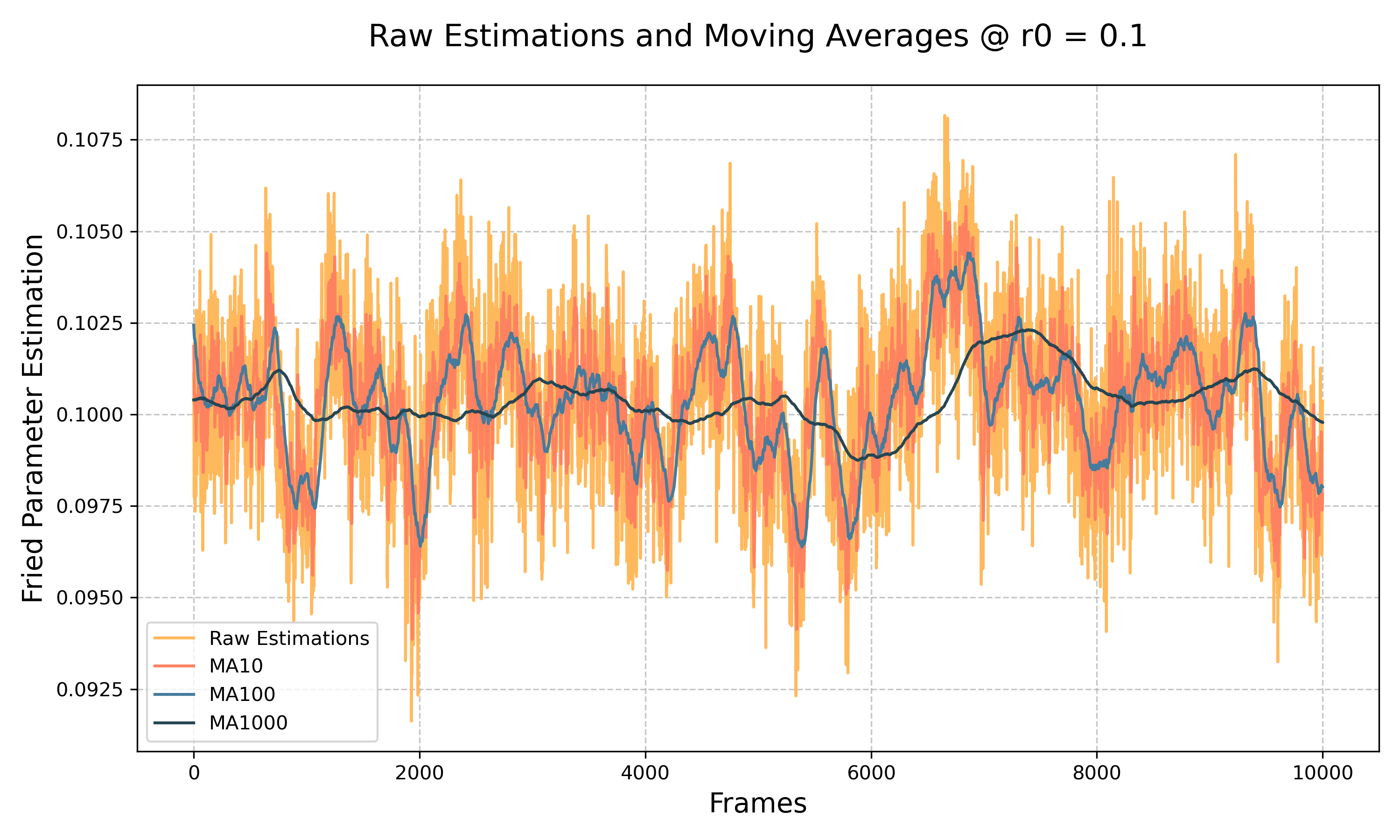}
\caption{Distribution of network estimations and their moving averages over 10, 000 consecutive frames at a constant r0 of 0.1m.}
\label{fig:SH_MA}
\end{figure}

To compute moving average measurements, we calculated the mean of network estimates over a specified number of consecutive frames extracted from the simulation data. In
this study, we employed a sliding window approach, considering different window sizes: [1, 10, 100, 1000]. We utilised the network trained with noise as outlined in {Section~\ref{Robust}} for these estimations and measured the variance of its predictions across a dataset comprising 10, 000 consecutive frames of closed-loop wavefront sensor images. Throughout this experiment, we maintained constant instrument and atmospheric configurations, including a guide star magnitude of 10 and the presence of both sources of noise. While we conducted this experiment across various r0 values, we focused on the $r0 = 0.1m$ case for our experimental analysis. The results are summarised in {Figure~\ref{fig:SH_MA}}. 

It is clear that the Fried parameter estimates vary from frame to frame, which indicates a small error in estimation. On inspection of the raw estimates in {Figure~\ref{fig:SH_MA}}, the variance of the estimation is clear and is reasonably consistent. The mean value shows that there is some other structure present which is a surprise and helps us understand the distributions we measured previously. In incumbent Fried parameter measurement methods, the process requires many second to minutes of data points, where Figure~\ref{fig:SH_MA} is a total of 10 seconds of inferred estimates so our single frame estimation method has a new level of detail in the temporal behaviour of the Fried parameter. We can speculate that there is some small variations in the Fried parameter involved with the simulated atmospheric model. This requires further investigation to understand and could be useful future work. 

Under consistent Fried parameter conditions, employing a moving average approach significantly diminishes estimation variance, leading to enhanced overall network stability. As anticipated, this enhanced stability relates directly to the size of the sliding window used for the moving average. 
%
%
These findings substantiate the practical utility of such techniques in real-world applications, offering a means to improve the stability and reliability of data-driven approaches for Fried parameter estimation.

In the real-world context, however, the Fried parameter does not remain static but can fluctuate due to evolving atmospheric conditions. Consequently, there may be scenarios where the application of moving average techniques could be less advantageous, especially in the presence of rapid atmospheric changes. In such cases, the moving average may lag in reflecting these quick alterations, posing a potential challenge to its effectiveness. 
These results affirm the viability of employing moving averages for enhancing
the stability of data-driven approaches, while recognizing their ideal applicability in
scenarios with relatively stable atmospheric conditions.

\subsection{Results - Pyramid Wavefront Sensor}

Motivated by the application of wavefront estimation from SH-WFS images~\cite{smith:2022, smith:2023a} to the Py-WFS~\cite{Pou:24}, the same methodology we have applied earlier to the SH-WFS is conducted with the pyramid WFS. As the pyramid WFS is a popular choice for new astronomical instrumentation, validating our methods for both sensors creates wider utility for our methods. For the sake of brevity, rather than repeat all of the results here we will show the what we consider the two key results for the pyramid WFS experiments. 

\subsubsection{Open and Closed-loop Estimators for the Py-WFS}

In previous {Sections~\ref{res:open}}{,~\ref{res:closed}}{,~\ref{openclosed}}, we discovered that our network-based methods could infer the Fried parameter from a single SH-WFS image in either open or closed-loop, and could be trained on both open and closed-loop SH-WFS images to infer accurate Fried parameter estimates. These same experiments were carried out for the Py-WFS, with a minor modification to the ANN structure to accommodate the four image segments of the Py-WFS. The results of this investigation of the Py-WFS are summarised in Table~\ref{tab:pyr_oc}.

\begin{table}[ht]
\caption{\textbf{Py-WFS analysis}. Summary of MAE and RMSE scores of open-loop, closed-loop and combined networks against open and closed-loop Py-WFS data across all $ r_{0} $ values.}
\label{tab:pyr_oc}
\centering
\begin{tabular}{ll|ll}
               &      & \multicolumn{2}{c}{\textbf{Evaluation Dataset}} \\
\textbf{Network} & (m)  & Open-loop            & Closed-loop           \\ \hline
Open-loop      & MAE  & 0.00319              & 0.03361               \\
               & RMSE & 0.00428              & 0.04120               \\ \hline
Closed-loop    & MAE  & 0.02279              & 0.00310               \\
               & RMSE & 0.02997              & 0.00402               \\ \hline
Combined       & MAE  & 0.00267              & 0.00258               \\
               & RMSE & 0.00351              & 0.00340              
\end{tabular}
\end{table}

Comparing the results for the pyramid WFS in Table~\ref{tab:pyr_oc} to the SH-WFS results in {Table~\ref{tab:sh_oc}}, it is clear we can also apply our Fried parameter estimation methods to the pyramid WFS. The overall accuracy appears to be slightly worse for the pyramid WFS, though this confirms again that the network trained on open and closed-loop sensor images is marginally more accurate than either the open or closed-loop data trained in isolation.

\subsubsection{Robustness of the Pyramid Wavefront Sensor to Photon Flux and Noise}\label{Robust_pyr}

The final network investigated for the SH-WFS in Section~\ref{Robust} was successfully trained and tested on closed-loop WFS data that included noise and several different guide star magnitudes. Here we conduct the same experiment for the Py-WFS, with the same data generation and training methods. The results of the robustness evaluation for the two WFS types are compared in Table~\ref{tab:pyr_gsmag_noise}.



\Omit{
\begin{table}[ht]
\caption{Summary of MAE and RMSE scores of open-loop, closed-loop and combined networks against open and closed-loop SH-WFS data across all $ r_{0} $ values. \label{tab:sh_oc}}
\centering
\begin{tabular}{ll|ll}
               &      & \multicolumn{2}{l}{\textbf{Evaluation Dataset}} \\
\textbf{Network} & (m)  & Open-loop             & Closed-loop             \\ \hline
Open-loop      & MAE  & 0.00184               & 0.00292                 \\
               & RMSE & 0.00242               & 0.00373                 \\ \hline
Closed-loop    & MAE  & 0.00682               & 0.00175                 \\
               & RMSE & 0.00906               & 0.00230                 \\ \hline
Combined       & MAE  & 0.00159               & 0.00159                 \\
               & RMSE & 0.00222               & 0.00221               
\end{tabular}
\end{table}
}
\Omit{
\begin{table}[ht]
\caption{Summary of MAE and RMSE scores of open-loop, closed-loop and combined networks against open and closed-loop Py-WFS data across all $ r_{0} $ values. \label{tab:pyr_oc}}
\centering
\begin{tabular}{ll|ll}
               &      & \multicolumn{2}{c}{\textbf{Evaluation Dataset}} \\
\textbf{Model} & (m)  & Open-loop            & Closed-loop           \\ \hline
Open-loop      & MAE  & 0.00319              & 0.03361               \\
               & RMSE & 0.00428              & 0.04120               \\ \hline
Closed-loop    & MAE  & 0.02279              & 0.00310               \\
               & RMSE & 0.02997              & 0.00402               \\ \hline
Combined       & MAE  & 0.00267              & 0.00258               \\
               & RMSE & 0.00351              & 0.00340              
\end{tabular}
\end{table}
}

\begin{table}[ht]
\caption{Summary of MAE and RMSE scores of the SH-WFS and Py-WFS models with noise, evaluated across closed-loop wavefront sensor images of magnitudes 9-13 with noise, and across all $ r_{0} $ values. Guide star magnitudes unseen during training are highlighted. \label{tab:pyr_gsmag_noise}}
\centering
\begin{tabular}{ll|lllll}
               &      & \multicolumn{5}{c}{\textbf{Guide Star Magnitude}} \\
\textbf{Network} & (m)  & 9        & 10       & 11      & 12      & 13      \\ \hline
SH-WFS         & MAE  & 0.00227  & 0.00218  & 0.00248 & \cellcolor{lightgray}0.01540 & \cellcolor{lightgray}0.03382 \\
               & RMSE & 0.00289  & 0.00278  & 0.00319 & \cellcolor{lightgray}0.02010 & \cellcolor{lightgray}0.04182 \\ \hline
Py-WFS         & MAE  & 0.00380  & 0.00384  & 0.00411 & \cellcolor{lightgray}0.07076 & \cellcolor{lightgray}0.08928 \\
               & RMSE & 0.00391  & 0.00391  & 0.00419 & \cellcolor{lightgray}0.07838 & \cellcolor{lightgray}0.09795
\end{tabular}
\end{table}

Similar to the previous {Section~\ref{Robust_pyr}}, we see that the pyramid WFS slightly under performs the SH-WFS for our network based methods. It is clear that the pyramid WFS is also robust to variations in photon flux and noise.

\Omit{
\begin{table}[ht]
\caption{Summary of MAE and RMSE scores of the SH-WFS and Py-WFS models with noise, evaluated across closed-loop wavefront sensor images of magnitudes 9-13 with noise, and across all $ r_{0} $ values. Guide star magnitudes unseen during training are highlighted. \label{tab:pyr_gsmag_noise}}
\centering
\begin{tabular}{ll|lllll}
               &      & \multicolumn{5}{c}{\textbf{Guide Star Magnitude}} \\
\textbf{Model} & (m)  & 9        & 10       & 11      & 12      & 13      \\ \hline
SH-WFS         & MAE  & 0.00227  & 0.00218  & 0.00248 & \cellcolor{gray}0.01540 & \cellcolor{gray}0.03382 \\
               & RMSE & 0.00289  & 0.00278  & 0.00319 & \cellcolor{gray}0.02010 & \cellcolor{gray}0.04182 \\ \hline
Py-WFS         & MAE  & 0.00380  & 0.00384  & 0.00411 & \cellcolor{gray}0.07076 & \cellcolor{gray}0.08928 \\
               & RMSE & 0.00391  & 0.00391  & 0.00419 & \cellcolor{gray}0.07838 & \cellcolor{gray}0.09795
\end{tabular}
\end{table}
}

\Omit{ 
\begin{figure}[ht]
  \includegraphics[width=\textwidth]{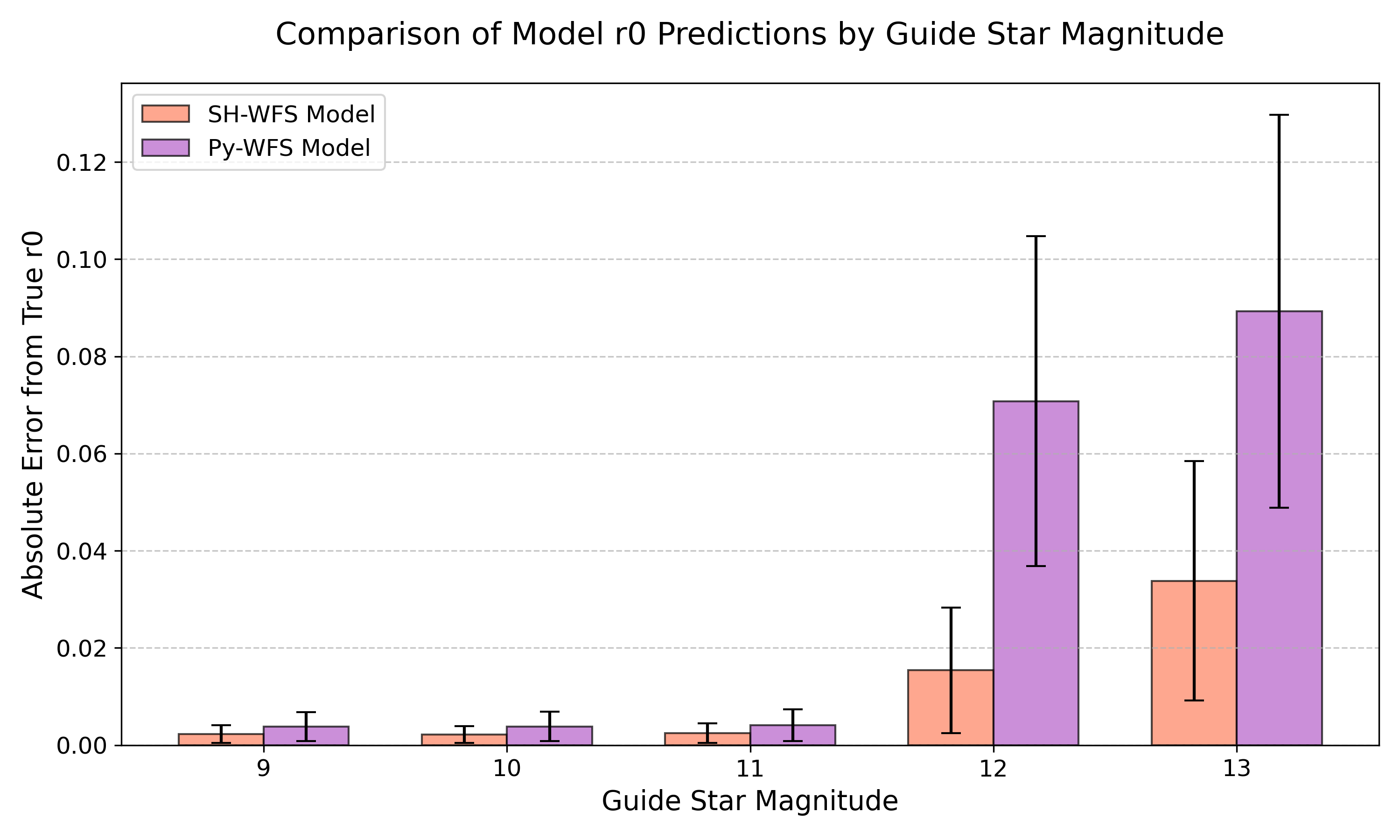}
  \caption{The absolute error of $ r_{0} $ predictions from the true $ r_{0} $ by guide star magnitude, compared across the SH-WFS model from Section~\ref{res:gsmag+noise} and Py-WFS model from Section~\ref{res:pyr_gsmag}. The error bars on the graph indicate one standard deviation.\label{fig:pyr_gsmag_noise}}
\end{figure}%
\clearpage
}
\section{Discussion and Future Work}
\label{sec:discussion}

This study presents a novel data-driven technique for real-time estimation of the Fried parameter in  directly from single WFS images in simulation. We achieve remarkable accuracy and robustness in the presence of simulated real-world parameters of variable photon flux and noise. The technique was validated for both open and closed-loop AO operations, revealing that a single network could be trained to estimate the Fried parameter from both open and closed-loop WFS images. 

In doing so, we found evidence that suggests our network-based method relied on features in the high spatial frequencies beyond the spatial cut-off frequency of the deformable mirror during AO operation to infer the Fried parameter. Our study further examined the ANN resilience to variations in guide star magnitudes and common noise sources, provided these variations remained within the bounds of its training data.

Performance evaluations using the SH-WFS highlighted an MAE of $0.23cm$ and an inference time of approximately $0.83ms$, all while running on modest off-the-shelf hardware. Furthermore, we illustrated the application of sliding window techniques to significantly improve the stability of real-time estimations, and the slight temporal wandering of the Fried parameter. Extending the analysis to the Py-WFS, the network achieved consistent results, underscoring its versatility across different wavefront sensor types and potential for broader applicability within various AO architectures. 

Unlike conventional approaches, our method operates independently of statistical models or specific instrument architectures, making it highly versatile and practical for AO applications. These findings have promising implications for advancing observation scheduling and AO controller optimisation, ultimately enhancing our capacity to obtain accurate turbulence information in the field of adaptive optics. 


While COMPASS simulations provided valuable datasets, their reliance on the Kolmogorov and von Kármán models limits their ability to fully replicate real-world atmospheric behaviour. Planned future research will focus on validating data-driven approaches with real-world on-sky data to assess practical applicability and evaluate the transferability of simulation-trained networks, and potentially training with on sky data.

Additionally, we plan to investigate the role of high spatial frequency features in inferring additional atmospheric parameters, aiming to deepen our understanding of the specific turbulence-related information retained in wavefront sensor images. This research could extend beyond parameter estimation and wavefront reconstruction to directly approximating turbulence profiles, potentially enabling novel wavefront sensors, applications and advancing fields that rely on atmospheric turbulence analysis.

\Omit{
\section{Template}
Begin the Introduction below the Keywords. The manuscript should not have headers, footers, or page numbers. It should be in a one-column format. References are often noted in the text and cited at the end of the paper.

\begin{table}[ht]
\caption{Fonts sizes to be used for various parts of the manuscript.  Table captions should be centered above the table.  When the caption is too long to fit on one line, it should be justified to the right and left margins of the body of the text.} 
\label{tab:fonts}
\begin{center}       
\begin{tabular}{|l|l|} 
\hline
\rule[-1ex]{0pt}{3.5ex}  Article title & 16 pt., bold, centered  \\
\hline
\rule[-1ex]{0pt}{3.5ex}  Author names and affiliations & 12 pt., normal, centered   \\
\hline
\rule[-1ex]{0pt}{3.5ex}  Keywords & 10 pt., normal, left justified   \\
\hline
\rule[-1ex]{0pt}{3.5ex}  Abstract Title & 11 pt., bold, centered   \\
\hline
\rule[-1ex]{0pt}{3.5ex}  Abstract body text & 10 pt., normal, justified   \\
\hline
\rule[-1ex]{0pt}{3.5ex}  Section heading & 11 pt., bold, centered (all caps)  \\
\hline
\rule[-1ex]{0pt}{3.5ex}  Subsection heading & 11 pt., bold, left justified  \\
\hline
\rule[-1ex]{0pt}{3.5ex}  Sub-subsection heading & 10 pt., bold, left justified  \\
\hline
\rule[-1ex]{0pt}{3.5ex}  Normal text & 10 pt., normal, justified  \\
\hline
\rule[-1ex]{0pt}{3.5ex}  Figure and table captions & \, 9 pt., normal \\
\hline
\rule[-1ex]{0pt}{3.5ex}  Footnote & \, 9 pt., normal \\
\hline 
\rule[-1ex]{0pt}{3.5ex}  Reference Heading & 11 pt., bold, centered   \\
\hline
\rule[-1ex]{0pt}{3.5ex}  Reference Listing & 10 pt., normal, justified   \\
\hline
\end{tabular}
\end{center}
\end{table} 

\begin{table}[ht]
\caption{Margins and print area specifications.} 
\label{tab:Paper Margins}
\begin{center}       
\begin{tabular}{|l|l|l|} 
\hline
\rule[-1ex]{0pt}{3.5ex}  Margin & A4 & Letter  \\
\hline
\rule[-1ex]{0pt}{3.5ex}  Top margin & 2.54 cm & 1.0 in.   \\
\hline
\rule[-1ex]{0pt}{3.5ex}  Bottom margin & 4.94 cm & 1.25 in.  \\
\hline
\rule[-1ex]{0pt}{3.5ex}  Left, right margin & 1.925 cm & .875 in.  \\
\hline
\rule[-1ex]{0pt}{3.5ex}  Printable area & 17.15 x 22.23 cm & 6.75 x 8.75 in.  \\
\hline 
\end{tabular}
\end{center}
\end{table}

LaTeX margins are related to the document's paper size. The paper size is by default set to USA letter paper. To format a document for A4 paper, the first line of this LaTeX source file should be changed to \verb|\documentclass[a4paper]{spie}|.   

Authors are encouraged to follow the principles of sound technical writing, as described in Refs.~\citenum{Alred03} and \citenum{Perelman97}, for example.  Many aspects of technical writing are addressed in the {\em AIP Style Manual}, published by the American Institute of Physics.  It is available on line at \url{https://publishing.aip.org/authors}. A spelling checker is helpful for finding misspelled words. 

An author may use this LaTeX source file as a template by substituting his/her own text in each field.  This document is not meant to be a complete guide on how to use LaTeX.  For that, please see the list of references at \url{http://latex-project.org/guides/} and for an online introduction to LaTeX please see \citenum{Lees-Miller-LaTeX-course-1}. 

\section{FORMATTING OF MANUSCRIPT COMPONENTS}

This section describes the normal structure of a manuscript and how each part should be handled.  The appropriate vertical spacing between various parts of this document is achieved in LaTeX through the proper use of defined constructs, such as \verb|\section{}|.  In LaTeX, paragraphs are separated by blank lines in the source file. 

At times it may be desired, for formatting reasons, to break a line without starting a new paragraph.  This situation may occur, for example, when formatting the article title, author information, or section headings.  Line breaks are inserted in LaTeX by entering \verb|\\| or \verb|\linebreak| in the LaTeX source file at the desired location.  

\subsection{Title and Author Information}
\label{sec:title}

The article title appears centered at the top of the first page.  The title font is 16 point, bold.  The rules for capitalizing the title are the same as for sentences; only the first word, proper nouns, and acronyms should be capitalized.  Avoid using acronyms in the title.  Keep in mind that people outside your area of expertise might read your article. At the first occurrence of an acronym, spell it out, followed by the acronym in parentheses, e.g., noise power spectrum (NPS). 

The author list is in 12-pt. regular, centered. Omit titles and degrees such as Dr., Prof., Ph.D., etc. The list of affiliations follows the author list. Each author's affiliation should be clearly noted. Superscripts may be used to identify the correspondence between the authors and their respective affiliations.  Further author information, such as e-mail address, complete postal address, and web-site location, may be provided in a footnote by using \verb|\authorinfo{}|, as demonstrated above.

\subsection{Abstract and Keywords}
The title and author information is immediately followed by the Abstract. The Abstract should concisely summarize the key findings of the paper.  It should consist of a single paragraph containing no more than 250 words.  The Abstract does not have a section number.  A list of up to eight keywords should immediately follow the Abstract after a blank line.  These keywords will be included in a searchable database at SPIE.

\subsection{Body of Paper}
The body of the paper consists of numbered sections that present the main findings.  These sections should be organized to best present the material.  See Sec.~\ref{sec:sections} for formatting instructions.

\subsection{Appendices}
Auxiliary material that is best left out of the main body of the paper, for example, derivations of equations, proofs of theorems, and details of algorithms, may be included in appendices.  Appendices are enumerated with uppercase Latin letters in alphabetic order, and appear just before the Acknowledgments and References. Appendix~\ref{sec:misc} contains more about formatting equations and theorems.

\subsection{Acknowledgments}
In the Acknowledgments section, appearing just before the References, the authors may credit others for their guidance or help.  Also, funding sources may be stated.  The Acknowledgments section does not have a section number.

\subsection{References}
SPIE is able to display the references section of your paper in the SPIE Digital Library, complete with links to referenced journal articles, proceedings papers, and books, when available. This added feature will bring more readers to your paper and improve the usefulness of the SPIE Digital Library for all researchers. The References section does not have a section number.  The references are numbered in the order in which they are cited.  Examples of the format to be followed are given at the end of this document.  

The reference list at the end of this document is created using BibTeX, which looks through the file {\ttfamily report.bib} for the entries cited in the LaTeX source file.  The format of the reference list is determined by the bibliography style file {\ttfamily spiebib.bst}, as specified in the \verb|\bibliographystyle{spiebib}| command.  Alternatively, the references may be directly formatted in the LaTeX source file.

For books\cite{Lamport94,Alred03,Goossens97}, the listing includes the list of authors, book title, publisher, city, page or chapter numbers, and year of publication.  A reference to a journal article\cite{Metropolis53} includes the author list, title of the article (in quotes), journal name (in italics, properly abbreviated), volume number (in bold), inclusive page numbers, and year.  By convention\cite{Lamport94}, article titles are capitalized as described in Sec.~\ref{sec:title}.  A reference to a proceedings paper or a chapter in an edited book\cite{Gull89a} includes the author list, title of the article (in quotes), volume or series title (in italics), volume number (in bold), if applicable, inclusive page numbers, publisher, city, and year.  References to an article in the SPIE Proceedings may include the conference name (in italics), as shown in Ref.~\citenum{Hanson93c}. For websites\cite{Lees-Miller-LaTeX-course-1} the listing includes the list of authors, title of the article (in quotes), website name, article date, website address either enclosed in chevron symbols ('\(<\)' and '\(>\)'),  underlined or linked, and the date the website was accessed. 

If you use this formatting, your references will link your manuscript to other research papers that are in the CrossRef system. Exact punctuation is required for the automated linking to be successful. 

Citations to the references are made using superscript numerals, as demonstrated in the above paragraph.  One may also directly refer to a reference within the text, e.g., ``as shown in Ref.~\citenum{Metropolis53} ...''

\subsection{Footnotes}
Footnotes\footnote{Footnotes are indicated as superscript symbols to avoid confusion with citations.} may be used to provide auxiliary information that doesn't need to appear in the text, e.g., to explain measurement units.  They should be used sparingly, however.  

Only nine footnote symbols are available in LaTeX. If you have more than nine footnotes, you will need to restart the sequence using the command  \verb|\footnote[1]{Your footnote text goes here.}|. If you don't, LaTeX will provide the error message {\ttfamily Counter too large.}, followed by the offending footnote command.

\section{SECTION FORMATTING}
\label{sec:sections}

Section headings are centered and formatted completely in uppercase 11-point bold font.  Sections should be numbered sequentially, starting with the first section after the Abstract.  The heading starts with the section number, followed by a period.  In LaTeX, a new section is created with the \verb|\section{}| command, which automatically numbers the sections.

Paragraphs that immediately follow a section heading are leading paragraphs and should not be indented, according to standard publishing style\cite{Lamport94}.  The same goes for leading paragraphs of subsections and sub-subsections.  Subsequent paragraphs are standard paragraphs, with 14-pt.\ (5 mm) indentation.  An extra half-line space should be inserted between paragraphs.  In LaTeX, this spacing is specified by the parameter \verb|\parskip|, which is set in {\ttfamily spie.cls}.  Indentation of the first line of a paragraph may be avoided by starting it with \verb|\noindent|.
 
\subsection{Subsection Attributes}

The subsection heading is left justified and set in 11-point, bold font.  Capitalization rules are the same as those for book titles.  The first word of a subsection heading is capitalized.  The remaining words are also capitalized, except for minor words with fewer than four letters, such as articles (a, an, and the), short prepositions (of, at, by, for, in, etc.), and short conjunctions (and, or, as, but, etc.).  Subsection numbers consist of the section number, followed by a period, and the subsection number within that section.  

\subsubsection{Sub-subsection attributes}
The sub-subsection heading is left justified and its font is 10 point, bold.  Capitalize as for sentences.  The first word of a sub-subsection heading is capitalised.  The rest of the heading is not capitalised, except for acronyms and proper names.  

\section{FIGURES AND TABLES}

Figures are numbered in the order of their first citation.  They should appear in numerical order and on or after the same page as their first reference in the text.  Alternatively, all figures may be placed at the end of the manuscript, that is, after the Reference section.  It is preferable to have figures appear at the top or bottom of the page.  Figures, along with their captions, should be separated from the main text by at least 0.2 in.\ or 5 mm.  

Figure captions are centered below the figure or graph.  Figure captions start with the figure number in 9-point bold font, followed by a period; the text is in 9-point normal font; for example, ``{\footnotesize{Figure 3.}  Original image...}''.  See Fig.~\ref{fig:example} for an example of a figure caption.  When the caption is too long to fit on one line, it should be justified to the right and left margins of the body of the text.  

Tables are handled identically to figures, except that their captions appear above the table. 

   \begin{figure} [ht]
   \begin{center}
   \begin{tabular}{c} 
   \includegraphics[height=5cm]{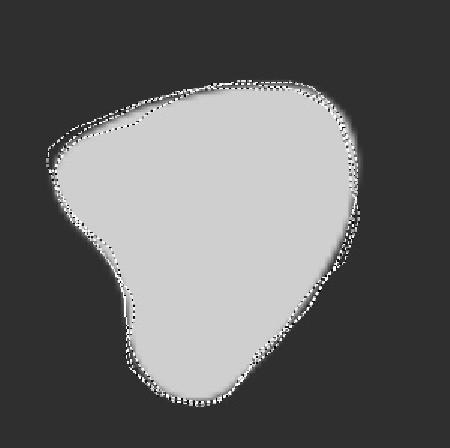}
   \end{tabular}
   \end{center}
   \caption[example] 
   { \label{fig:example} 
Figure captions are used to describe the figure and help the reader understand it's significance.  The caption should be centered underneath the figure and set in 9-point font.  It is preferable for figures and tables to be placed at the top or bottom of the page. LaTeX tends to adhere to this standard.}
   \end{figure} 

\section{MULTIMEDIA FIGURES - VIDEO AND AUDIO FILES}

Video and audio files can be included for publication. See Tab.~\ref{tab:Multimedia-Specifications} for the specifications for the mulitimedia files. Use a screenshot or another .jpg illustration for placement in the text. Use the file name to begin the caption. The text of the caption must end with the text ``http://dx.doi.org/doi.number.goes.here'' which tells the SPIE editor where to insert the hyperlink in the digital version of the manuscript. 

Here is a sample illustration and caption for a multimedia file:

   \begin{figure} [ht]
   \begin{center}
   \begin{tabular}{c} 
   \includegraphics[height=5cm]{MultimediaFigure.jpg}
	\end{tabular}
	\end{center}
   \caption[example] 
   { \label{fig:video-example} 
A label of “Video/Audio 1, 2, …” should appear at the beginning of the caption to indicate to which multimedia file it is linked . Include this text at the end of the caption: \url{http://dx.doi.org/doi.number.goes.here}}
   \end{figure} 
   
   \begin{table}[ht]
\caption{Information on video and audio files that must accompany a manuscript submission.} 
\label{tab:Multimedia-Specifications}
\begin{center}       
\begin{tabular}{|l|l|l|}
\hline
\rule[-1ex]{0pt}{3.5ex}  Item & Video & Audio  \\
\hline
\rule[-1ex]{0pt}{3.5ex}  File name & Video1, video2... & Audio1, audio2...   \\
\hline
\rule[-1ex]{0pt}{3.5ex}  Number of files & 0-10 & 0-10  \\
\hline
\rule[-1ex]{0pt}{3.5ex}  Size of each file & 5 MB & 5 MB  \\
\hline
\rule[-1ex]{0pt}{3.5ex}  File types accepted & .mpeg, .mov (Quicktime), .wmv (Windows Media Player) & .wav, .mp3  \\
\hline 
\end{tabular}
\end{center}
\end{table}

\appendix    

\section{MISCELLANEOUS FORMATTING DETAILS}
\label{sec:misc}

It is often useful to refer back (or forward) to other sections in the article.  Such references are made by section number.  When a section reference starts a sentence, Section is spelled out; otherwise use its abbreviation, for example, ``In Sec.~2 we showed...'' or ``Section~2.1 contained a description...''.  References to figures, tables, and theorems are handled the same way.

\subsection{Formatting Equations}
Equations may appear in line with the text, if they are simple, short, and not of major importance; e.g., $\beta = b/r$.  Important equations appear on their own line.  Such equations are centered.  For example, ``The expression for the field of view is
\begin{equation}
\label{eq:fov}
2 a = \frac{(b + 1)}{3c} \, ,
\end{equation}
where $a$ is the ...'' Principal equations are numbered, with the equation number placed within parentheses and right justified.  

Equations are considered to be part of a sentence and should be punctuated accordingly. In the above example, a comma follows the equation because the next line is a subordinate clause.  If the equation ends the sentence, a period should follow the equation.  The line following an equation should not be indented unless it is meant to start a new paragraph.  Indentation after an equation is avoided in LaTeX by not leaving a blank line between the equation and the subsequent text.

References to equations include the equation number in parentheses, for example, ``Equation~(\ref{eq:fov}) shows ...'' or ``Combining Eqs.~(2) and (3), we obtain...''  Using a tilde in the LaTeX source file between two characters avoids unwanted line breaks.

\subsection{Formatting Theorems}

To include theorems in a formal way, the theorem identification should appear in a 10-point, bold font, left justified and followed by a period.  The text of the theorem continues on the same line in normal, 10-point font.  For example, 

\noindent\textbf{Theorem 1.} For any unbiased estimator...

Formal statements of lemmas and algorithms receive a similar treatment.

}

\Omit{ 
\acknowledgments 
This paper is a summarized report of a deep investigation by Taisei Fujii conducted for an honors thesis in computer science at the Australian National University in 2023. Due to format limitations, some areas of this paper are necessarily compressed. The original work and code is available by request.
  }

\bibliography{report} 
\bibliographystyle{spiebib} 

\end{document}